\documentclass[10pt, conference, letterpaper]{IEEEtran}

\usepackage{verbatim,epsfig,times,subfigure,setspace}
\usepackage{amsmath,amsfonts,amssymb,mathrsfs}
\usepackage{enumerate,multirow,indentfirst}
\usepackage{bm,bbm}
\usepackage[square, comma, sort&compress, numbers]{natbib}
\usepackage[T1]{fontenc}
\usepackage{url}

\usepackage{subfigure}
\usepackage{float}

\begin{document}

% environment
\newtheorem{assumption}{\bf Assumption}
\newtheorem{definition}{\bf Definition}
\newtheorem{theorem}{\bf Theorem}
\newtheorem{lemma}{\bf Lemma}
\newtheorem{corollary}{\bf Corollary}

\title{\LARGE{On Optimal Service Differentiation in Congested Network Markets}}

\author{Mao Zou$^{1,2}$, Richard T. B. Ma$^3$, Xin Wang$^{1,2}$, Yinlong Xu$^{1,2}$ \\
$^1$ School of Computer Science and Technology, University of Science and Technology of China\\
$^2$ AnHui Province Key Laboratory of High Performance Computing (USTC)\\
$^3$ School of Computing, National University of Singapore\\
{raffaele@mail.ustc.edu.cn, tbma@comp.nus.edu.sg, yixinxa@mail.ustc.edu.cn, ylxu@ustc.edu.cn}
}

\maketitle
\begin{abstract}
As Internet applications have become more diverse in recent years, 		
%In recent years the diversities of Internet applications and users have risen dramatically.
users having heavy demand for online video services are more willing to pay higher prices for better services than light users that mainly use e-mails and instant messages. This encourages the Internet Service Providers (ISPs) to explore service differentiations so as to optimize their profits and allocation of network resources. Much prior work has focused on the viability of network service differentiation by comparing with the case of a single-class service. However, the optimal service differentiation for an ISP subject to resource constraints has remained unsolved. In this work, we establish an optimal control framework to derive the analytical solution to an ISP's optimal service differentiation, i.e. the optimal service qualities and associated prices. By analyzing the structures of the solution, we reveal how an ISP should adjust the service qualities and prices in order to meet varying capacity constraints and users' characteristics. We also obtain the conditions under which ISPs have strong incentives to implement service differentiation and whether regulators should encourage such practices.
\end{abstract}

%
%  Use this command to print the description
%

\section{Introduction}
Recent years have witnessed increasing commercialization of the Internet and considerably rising diversities of Internet applications and users. Users having heavy demand for quality-sensitive services such as online video services are often willing to pay higher prices for better service qualities, while light users tend to have low willingness-to-pay for elementary services such as e-mails and instant messages. Such heterogeneity in the users' characteristics has rendered the traditional practice of uniform prices, regardless of the quality of service, inefficient in both generating profits and allocating network capacity. In light of this, there has been tremendous interest in designing pricing strategies to provide differentiated services. Many schemes \cite{odlyzko1999paris,odlyzko2004pricing,odlyzko2004evolution} have been designed, e.g., \textit{Paris Metro Pricing} (PMP)  \cite{odlyzko1999paris} proposed by Andrew Odlyzko. In such a scheme, network is partitioned into several logically separated channels, each of which uses a fraction of the entire network capacity and a different price, and the channels with higher prices attract fewer users but  provide better services.

An important theoretical question naturally arises: what is the optimal service differentiation for an ISP subject to its capacity constraints so as to maximize its profit? Despite the vast literature on network service differentiation, this question has remained unanswered. Prior work \cite{chau2010viability,shakkottai2008price,ma2014pay,jain2001analysis} mostly focused on the viability of PMP-type of differentiation and did not investigate the optimal service differentiation. 
%In order to understand the major challenges 
To tackle the optimal service differentiation, we delve a bit into the ISP's decision-making process. In deciding the optimal strategy, an ISP needs to answer two questions: what are the service qualities to offer and what are the associated prices for them. For the former question, even the range of the optimal set of service qualities is hard to determine; for the latter one, it needs to find the optimal quality-price mapping for the chosen service qualities. The problem is actually a dynamic optimization problem in a functional space and the domain of the price mapping is to be determined. 
As the realized quality in a service class depends on the allocated network capacity and the aggregate user demand, which is determined by the available service qualities and the corresponding prices,
another major challenge is to model the ISP's capacity allocation plan subject to network capacity constraints and determine the realized quality in terms of network congestion. In other words, ISPs need to allocate limited network capacity to each service class so as to guarantee the promised service quality. 
%while the quality also depends on the aggregate user demand which in turn is determined by the range of service qualities and the corresponding prices.

In view of the complexity of the optimal service differentiation, the common practice of prior work is to adopt simplifying assumptions. For instance, Jain \textit{et al.}~\cite{jain2001analysis} assumed the ISP only provides two service classes.  Shakkottai \textit{et al.}~\cite{shakkottai2008price} overlooked the congestion externalities and considered a loose upper bound (under the first degree price discrimination) of the ISP's optimal profits instead. Chau \textit{et al.}~\cite{chau2010viability} considered a finite number of service classes, analyzed the viability of differentiated pricing, but did not solve the optimal strategy.

In this work, we solve an ISP's optimal service differentiation in generic settings by explicitly modelling the ISP's capacity allocation plan subject to network capacity constraints and incorporating the dual approach in microeconomics \cite{mirrlees1971exploration} to transform the problem into a tractable optimal control problem. As for the network capacity constraints, we consider two typical scenarios: \textit{a)} the fixed capacity scenario, where an ISP has limited network capacity, and \textit{b)} the variable capacity scenario, where an ISP can invest to expand the capacity of its network infrastructure. Our findings include the following.
\begin{itemize}
\item We formulate a profit maximization problem subject to capacity constraints under an optimal control framework
and derive the analytical optimal service differentiation  of an ISP, i.e., the qualities and  prices of service classes.
\item We find that the optimal differentiation strategy offers service classes with sufficiently high qualities and prices such that users with low value will not join the services, even under abundant capacity. We characterize an ISP's market share by users' virtual valuation function \cite{myerson1981optimal}.
\item We show that when an ISP expands its capacity, it should increase  market share by offering more premium service classes and reducing prices.  When there are more high-end users, it is optimal for the ISP to offer a narrower range of premium service classes with higher prices.
\item We show that if network capacity is scarce, an ISP has strong incentives to implement service differentiation and regulators should encourage such practices in order to improve the total user surplus; otherwise, it is optimal for the ISP to provide a single-class service.
\end{itemize}

To the best of our knowledge, our work is the first to analytically derive and characterize an ISP's optimal service differentiation subject to network capacity constraints. We believe that our findings will help  ISPs to better price  network services and guide regulators to design regulations.
The rest of the paper is organized as follows. Section II discusses related work. Section III presents the model and formulate the ISP's profit-maximizing problem as a nested optimization problem. In Section IV, we derive the analytical solution of the optimal service differentiation and study the structure of the solution. We reveal the dynamics of the solution under varying market environments in Section V and conclude in Section VI.

\section{Related Work}
Much theoretical economics literature has considered product differentiation. Mussa and Rosen \cite{mussa1978monopoly} analyzed the optimal product mix for a monopolist offering several qualities of a product to consumers with different preferences for quality. Maskin and Riley \cite{maskin1984monopoly} characterized the optimal selling strategy of a monopolist, and Champsaur and Rochet \cite{champsaur1989multiproduct} studied the product differentiation for two competing firms. Both adopt a common assumption that the quality of the product can be unilaterally decided by the seller. However, these models are inapplicable to network services because the quality of services are influenced by both the user demand and the network capacity. Chander and Leruth \cite{chander1989optimal} studied the best strategy of a monopolist in the presence of congestion effects. Nevertheless, they did not provide the solution of the optimal strategy and their congestion model does not capture the  scarcity of capacity in network markets.

Various pricing schemes for providing differentiated network services has been proposed \cite{odlyzko1999paris,odlyzko2004pricing,odlyzko2004evolution}. In particular, Paris Metro Pricing (PMP) \cite{odlyzko1999paris} has attracted considerable attention for its simplicity. Since PMP does not guarantee the quality of service for users, which depends on the spontaneous user demand, its viability and effectiveness need to be established. Prior studies \cite{chau2010viability,shakkottai2008price,ma2014pay,jain2001analysis,gibbens2000internet} addressed the viability of PMP. Shakkottai \textit{et al.} \cite{shakkottai2008price} characterized the types of environments in which simple flat-rate pricing is efficient in extracting profits. Lee \textit{et al.} \cite{lee2012price} extended the model of \cite{shakkottai2008price} to explicitly consider the congestion externalities and indicate that when the network is congested, a simple flat-rate pricing is inefficient in extracting profits. Jain \textit{et al.} \cite{jain2001analysis} analyzed the ISP's profit-maximizing problem when the ISP is restricted to provider two service classes. Chau \textit{et al.} \cite{chau2010viability} provided sufficient condition of congestion externalities for the viability of PMP. Several studies investigated the competition among service providers. For example, Gibbens \textit{et al.} \cite{gibbens2000internet} analyzed the competition between two ISP's and show that in any equilibrium competitive outcome, both ISPs offer a single-class service.  Ma \cite{ma2014pay} studied usage-based pricing scheme and competition among oligopolistic service provides. These studies focused on the viability of PMP and its variants by comparing with a single-class service. Our work, however, focuses on the theoretical problem of solving the ISP's optimal service differentiation subject to capacity constraints. We establish an optimal control framework to derive the analytical solution to the ISP's optimal service differentiation and characterize its properties.

\section{Model}
\subsection{ISP and User Model}
We consider an ISP and a continuum of users with heterogeneous characteristics. Suppose the ISP offers a set $\mathcal{N}$ of service classes and adopts flat-rate\footnote {Our analysis can be extended to handle usage-based pricing schemes. Interested readers are referred to the Appendix for details.} pricing \cite{anania1997flat} for all service classes. For each service class $i \in \mathcal{N}$, we denote $p_{i}$ as the price and $q_{i}$ as the congestion level, a measure of service quality. Correspondingly, we denote $\mathbf{p} \triangleq \{p_{i} : i\in\mathcal{N}\}$ and $\mathbf{q} \triangleq \{q_{i} : i\in\mathcal{N}\}$ as the price and congestion vectors.

Once the menu of services $(\mathbf{p},\mathbf{q})$ is decided, users will make choices according to their preferences over the bundles $\{(p_{i},q_{i}) : i\in\mathcal{N}\}$. We characterize each user by her intrinsic value of the network service $\theta\in\Theta=[0,1]$.
We denote  $f(\theta)$ and $F(\theta)$ as the probability density and cumulative distribution functions of users' value, and assume that $f(\theta)$ is continuously differentiable and strictly positive over $\Theta$. %We denote $F(\theta)$ as the corresponding cumulative distribution function.
When a user with value $\theta$ chooses a service class $(p,q)$, we define her utility as
\begin{equation}\label{utility}
u(p,q;\theta)=\theta v(q)-p,
\end{equation}
which can be interpreted as follows.
\begin{itemize}
\item $v(\cdot)$ is a satisfaction discount function which captures the negative effect of congestion on users' intrinsic values.
\item $\theta v(q)$ is the user's achieved value over the service under a congestion level $q$, and thus $u(p,q;\theta)$ is the user's net surplus, i.e., the achieved value $\theta v(q)$ minus the price $p$.
\item Without loss of generality, we normalize the range of congestion level q to the interval $[0,1]$ and assume that $v(\cdot): [0,1] \rightarrow \mathbb{R}_{+}$ is continuously differentiable and strictly decreasing, satisfying $v(0)=1$ and $v(1)=0$. This models that the user's value is at maximum without congestion, i.e., $q=0$, and decreases to zero when the network is heavily congested, i.e., $q\rightarrow 1$.
\end{itemize}

We assume that users are \textit{individually rational}, i.e., any user with value $\theta$ chooses either \textit{a)} to join the service class $k$ that yields the highest nonnegative utility\footnote{Technically, we should write $k\in\arg\max_{j\in \mathcal{N}}\;u(p_j,q_j;\theta)$. By using the equal sign, we assume when there are more than one service classes that maximize the user's utility, the user will choose the one by her own criteria.}:
\begin{equation}\label{user_choice}
k=\arg\max_{j\in \mathcal{N}}\;u(p_j,q_j;\theta)
\end{equation}
or \textit{b)} to opt out of the ISP's services if joining any service class induces a negative utility, i.e., \(u(p_{i},q_{i};\theta)<0, \forall i \in \mathcal{N}\).
To unify the above two cases, we assume the existence of a \textit{dummy service class} $0$ which satisfies $p_{0}=0$ and $q_{0}=1$, and therefore, any user can always choose service class $0$ to gain zero utility.
% and there will be no need to differentiate between the above two cases.
We denote the set of user types that choose the service class $i$ as \(\Theta_{i}(\mathbf{p},\mathbf{q})\).
Notice that if the ISP offers two service classes $i$ and $j$ such that $p_i>p_j$ and $q_i>q_j$, no user would choose  class $i$ because it is inferior to class $j$ in terms of both price and congestion. Therefore, without loss of generality, we can sort the indexes of non-dummy service classes in an ascending order of congestion level as follows:
\begin{displaymath}
p_1>p_2>...>p_{|\mathcal{N}|} \quad\text{and}\quad q_1<q_2<...<q_{|\mathcal{N}|}.
\end{displaymath}

\begin{lemma}\label{delimeter}
Given that $|\mathcal{N}|<+\infty$, we define a set of vectors $\big\{\theta_{i}|i=0,1,...,|\mathcal{N}|\big\}$ as
\begin{align}
\begin{cases}
\displaystyle\theta_0=1,\ \theta_{|\mathcal{N}|}=\frac{p_{|\mathcal{N}|}}{v(q_{|\mathcal{N}|})},\\
\displaystyle\theta_i=\frac{p_i-p_{i+1}}{v(q_i)-v(q_{i+1})}, \ \forall i=1,2,...,|\mathcal{N}|-1.\\
\end{cases}
\end{align}
If $1\geq \theta_1 \geq ... \geq \theta_{|\mathcal{N}|}$, then $\Theta_{i}(\mathbf{p},\mathbf{q})$ can be characterized by
\begin{equation}
\Theta_{i}(\mathbf{p},\mathbf{q})=[\theta_{i},\theta_{i-1}], \quad i=1,2,...,|\mathcal{N}|.
\end{equation}
\end{lemma}

%\begin{IEEEproof}[\bf Proof of Lemma \ref{delimeter}]
%\end{IEEEproof}

Lemma \ref{delimeter}\footnote{Due to space limitation, we defer all the proofs to the Appendix.} states that if each service class can capture some users, the set of user types captured by the service class $i$ is an interval $[\theta_{i},\theta_{i-1}]$. If there exists any service class that captures no users, we can always exclude it from the ISP's menu of services without affecting its profits and capacity consumption. Thus, without loss of generality, we assume that the ISP does not offer such empty service classes and the condition of Lemma \ref{delimeter} naturally holds, and consequently, the population of users that choose service class $i$ can be calculated as
\begin{equation}\label{demand}
d_{i}(\mathbf{p},\mathbf{q})\triangleq \int_{\Theta_{i}(\mathbf{p},\mathbf{q})} f(\theta) \mathrm{d} \theta = \int_{\theta_{i}}^{\theta_{i-1}}f(\theta)\mathrm{d}\theta,
\end{equation}
and the ISP's revenue can be expressed as $\sum_{i \in \mathcal{N}}p_{i}d_{i}(\mathbf{p},\mathbf{q})$.

\subsection{Network Congestion and Capacity Allocation}
Since the ISP usually has limited network capacity, it needs to design a capacity allocation plan to guarantee the promised service quality in terms of congestion in each service class. In general, the required amount of network capacity by a service class is determined by its promised service congestion and user population. For each service class $i \in \mathcal{N}$, we model the implied capacity to guarantee a congestion level $q_{i}$ under the aggregate user population $d_{i}$ as a function $C(d_{i},q_{i})$, which decreases in $q_i$ and increases in $d_i$.

\begin{assumption}\label{implied_c}
The implied capacity function satisfies $C(d,q) \triangleq d\cdot w(q)$, where $w(q): [0,1]\rightarrow\mathbb{R}_{+}$ is continuously differentiable, decreasing in $q$ and satisfies $w(1)=0$.
\end{assumption}

The form of implied capacity function in Assumption \ref{implied_c} models the \textit{capacity sharing} \cite{chau2010viability,gibbens2000internet} nature of network services, under which the service capacity is shared by all users. The function $w(q)$ models the capacity required to maintain the congestion level $q$ given that the service class has a unit user demand. Under Assumption \ref{implied_c}, the implied capacity function satisfies $C(\epsilon d,q)=\epsilon C(d,q)$, indicating that users will not perceive any difference in terms of congestion after service partitioning or multiplexing. This form is a generalization of $C(d,q)=d/q$ (to see this, let $w(q)=1/q$), which has been considered in much prior work \cite{chau2010viability,gibbens2000internet,jain2001analysis,ma2014pay}.

We model the ISP's capacity allocation plan by a vector $\mathbf{c} \triangleq \{c_i : \forall i \in \mathcal{N}\}$, where $c_i$ is the capacity allocated to service class $i$. To guarantee the congestion level of each service class, the ISP's capacity allocation plan $\mathbf{c}$ must satisfy
\begin{equation}\label{cap_constraint}
c_{i}\geq C(d_{i}(\mathbf{p},\mathbf{q}),q_{i}), \quad \forall i\in\mathcal{N}.
\end{equation}
Intuitively, when the ISP has limited network capacity and some of the inequalities in (\ref{cap_constraint}) do not bind, it probably has not taken full advantage of the capacity and can gain more profits by readjusting the menu of services. We will strictly characterize this in Theorem \ref{fixedc_sol} in Section IV-C.

We have by far established the relationship between the ISP's menu \((\mathbf{p},\mathbf{q})\) of services and the capacity allocation plan $\mathbf{c}$. We next proceed to analyze the ISP's limitation in capacity and formulate the ISP's profit-maximizing problem.

\subsection{Profit Maximization Problem}
We first analyze the network capacity constraints of the ISP. In this work, we focus on two typical scenarios:
\begin{itemize}
\item (Fixed Capacity) The ISP's services are supported by an existing network infrastructure and hence face a maximum network capacity, denoted by $C_M$.
\item (Variable Capacity) The ISP may invest to expand its network infrastructure. In this scenario, the ISP needs to find out the optimal amount of capacity to consume (invest). We capture the cost of consuming capacity $c$ in total by $S(c)$, a continuously nondecreasing function.
\end{itemize}

The ISP aims to find the optimal service menu $(\mathbf{p},\mathbf{q})$ and the capacity allocation plan $\mathbf{c}$ subject to the capacity constraints in order to maximize its profits. In the fixed capacity scenario, the allocated capacity should not exceed the maximum capacity, i.e., \(\textstyle\sum_{i \in \mathcal{N}} C\big(d_{i}(\mathbf{p},\mathbf{q}),q_{i}\big)\leq C_M\), and the ISP's profit maximization problem is formulated as follows.
\begin{align}\label{fixed_cap}
\underset {\mathcal{N},\mathbf{p},\mathbf{q}}{\text{maximize}}\quad &J = \sum_{i \in \mathcal{N}}p_{i}d_{i}(\mathbf{p},\mathbf{q})\\
\text{subject to}\quad &\Theta_{i}(\mathbf{p},\mathbf{q}) = \{\theta \in \Theta : i = \arg\max_{j \in \mathcal{N}} \  \theta v(q_{j})-p_{j}\},\nonumber\\
&d_{i}(\mathbf{p},\mathbf{q}) =
\begin{matrix}\int_{\Theta_{i}(\mathbf{p},\mathbf{q})} f(\theta) \mathrm{d}\theta\end{matrix}
, \ \forall i \in \mathcal{N},\nonumber\\
& \text{and} \quad \textstyle\sum_{i \in \mathcal{N}} C\big(d_{i}(\mathbf{p},\mathbf{q}),q_{i}\big)\leq C_M.\nonumber
\end{align}
Likewise, in the variable capacity scenario, the ISP's investment in capacity is characterized by
$S(\sum_{i \in \mathcal{N}}C(d_{i}(\mathbf{p},\mathbf{q}),q_{i}))$, and the profit maximization problem is formulated as follows.

\begin{align}\label{capexpansion}
\underset {\mathcal{N},\mathbf{p},\mathbf{q}}{\text{maximize}}\quad &J \!= \!\!\sum_{i \in \mathcal{N}}p_{i}d_{i}(\mathbf{p},\mathbf{q})
 \!-\! \!S\bigg(\!\sum_{i \in \mathcal{N}}C\big(d_{i}(\mathbf{p},\mathbf{q}),q_{i}\big)\!\!\bigg)\\
 \text{subject to}\quad & \Theta_{i}(\mathbf{p},\mathbf{q}) = \{\theta \in \Theta :
 i \!= \!\arg\max_{j \in \mathcal{N}} \  \theta v(q)\!-\!p(q)\}\nonumber\\
 & \text{and} \quad d_{i}(\mathbf{p},\mathbf{q}) =
 \begin{matrix}\int_{\Theta_{i}(\mathbf{p},\mathbf{q})} f(\theta) \mathrm{d}\theta\end{matrix}
 , \quad\forall i \in \mathcal{N}.\nonumber
\end{align}
By far, we have formulated the ISP's profit maximization problems as nested optimization problems with equilibrium constraints. In the next section, we demonstrate how they can be transformed into more tractable optimal control problems by using a dual approach and derive the analytic solutions.

\section{Optimal Control Framework}

Two of the main challenges in solving the nested optimization problems (\ref{fixed_cap}) and (\ref{capexpansion}) include: 1) the set of optimal service classes $\mathcal{N}^*$ is hard to predetermine, and 2)
%when the ISP decide the set of service classes $\mathcal{N}$ to offer, a natural question arises: whether it is optimal to offer a finite set of service classes, an interval of service classes, or even multiple disjoint intervals;
the sets of user demand $\Theta_{i}(\mathbf{p},\mathbf{q})$ are subject to the nonlinear user utility maximization problems which require different treatments depending on the connectedness of $\mathcal{N}$.
To address these problems, we need a unified and succinct representation of the ISP's strategy.
This section is organized as follows. Subsection A introduces a dual representation of the ISP's strategy and subsection B reformulates the ISP's profit maximization problem into a tractable optimal control problem using a dual approach. In subsection C, we presents the analytic solution to the ISP's optimal service differentiation. In subsection D, we study the characteristics of the optimal solution.

\subsection{Dual Approach}
We denote $\mathcal{Q} \subseteq [0,1]$ as a set of provided service qualities in terms of congestion and the ISP's pricing and capacity allocation plan can be represented by two mappings $p(\cdot): \mathcal{Q} \rightarrow \mathbb{R}_+$ and $c(\cdot): \mathcal{Q} \rightarrow \mathbb{R}_+$, which map levels of congestion to the corresponding capacities and prices. As mentioned before, we assume that $\mathcal{Q}$ contains a \textit{dummy service class} with congestion level $q=1$ and $p(q)=c(q)=0$.
Since the optimal service domain $\mathcal{Q}^*$ of the mappings is also undetermined, the strategy space is difficult to handle. To this end, we introduce a dual approach,  widely used in economics \cite{mirrlees1971exploration,champsaur1989multiproduct}, to describe the ISP's decision in terms of the user's choice, rather than the levels of congestion and prices.

\begin{definition}\label{dual} For any type $\theta$, we define $q(\theta)$ as the chosen service quality that maximizes the user's utility, satisfying
%	 must subscribe to a service class in the set
\begin{equation}
q(\theta) \in \arg\max_{q\in\mathcal{Q}}\theta v(q)-p(q).
\end{equation}
%i.e., the choice of user $\theta$ is given by \(q(\theta) \in X(\theta)\).
The \textit{indirect utility} of user type $\theta$ is defined by
\begin{equation}
V(\theta) \triangleq  \max_{q\in\mathcal{Q}} \  \theta v(q)-p(q).
\end{equation}
\end{definition}

When there exist multiple optimal choices, we assume that user $\theta$ chooses one from them according to her own criteria. \footnote{As we explain in the Appendix, this will not affect the correctness of later analysis.} Given a finite set (or more generally a compact set)  of choices, the optimal choice $q(\theta)$ always exists. The indirect utility $V(\theta)$ defines the maximum utility that can be achieved by users of type $\theta$ under their optimal choice of services.
The advantage of the dual approach is that the ISP's strategy $(\mathcal{Q},p(\cdot))$ can be equivalently represented by the user's indirect utility $V(\cdot)$ and optimal choice $q(\cdot)$, both of which have favorable properties and a fixed domain, i.e.,  $\Theta=[0,1]$.
Next, we characterize the properties of $q(\cdot)$ and $V(\cdot)$. %We ignore the peculiar case in which the user's utility maximization problem has no solution and assume that $X(\theta)\neq \emptyset$ for all $\theta \in \Theta$.
\begin{lemma}\label{monotonicity}
The user's optimal choice function $q(\cdot): [0,1]\rightarrow\mathcal{Q}$ is nonincreasing in $\theta$.
\end{lemma}

The Lemma \ref{monotonicity} states that a user with higher value would not choose a service class with worse congestion, and whatever the user's criteria of breaking ties of optimal choices is, the monotonicity of $q(\cdot)$ always holds. An immediate consequence is that $q(\cdot)$ is piecewise continuous.

By Definition \ref{dual}, any user of type $\theta$'s maximum utility $V(\theta)$ can also be expressed by the following equation:
\begin{equation}
V(\theta)=\theta v(q(\theta))-p(q(\theta)), \quad \forall \theta \in \Theta.
\end{equation}
Furthermore, according to the envelope theorem \cite{milgrom2002envelope}, $V(\cdot)$ is absolutely continuous and hence differentiable almost everywhere, and thus we have
\begin{equation}
V'(\theta)=v(q(\theta)), \ \ \text{for almost all} \ \  \theta \in \Theta.
\end{equation}
%And the equivalent integral form:
%\begin{equation}\label{u_integral}
%V(\theta)=V(0)+\int\nolimits_{0}^{\theta}v(q(s))\mathrm{d}s
%\end{equation}
The following lemma allows us to represent an ISP's pricing strategy using $V(\cdot)$ and $q(\cdot)$.
\begin{lemma}\label{dual_approach}
Let $q(\cdot)$ and $V(\cdot)$ be any two mappings from $\Theta$ to $\mathbb{R}$ and $\mathcal{Q}$ the range of $q(\cdot)$. If $q(\cdot)$ is nonincreasing and
\begin{displaymath}V(\theta)=V(0)+\int\nolimits_{0}^{\theta}v(q(s))\mathrm{d}s, \ \ \forall \theta \in \Theta,\end{displaymath}
then there exists an unique pricing $p(\cdot): \mathbb{R}_{+} \rightarrow \mathbb{R}$ such that
\begin{displaymath}
V(\theta) = \max_{q \in \mathcal{Q}} \ \theta v(q)-p(q) \ \text{and} \ q(\theta) \in \arg\max_{q \in \mathcal{Q}} \ \theta v(q)-p(q).
\end{displaymath}
\end{lemma}

By the envelope theorem \cite{milgrom2002envelope}, the indirect utility function $V(\cdot)$ and the user's choice function $q(\cdot)$ meet the condition in Lemma \ref{dual_approach}. Therefore, given the user's choice function $q(\cdot)$  and an initial level $V(0)$ of the indirect utility, we can recover the ISP's service differentiation strategy as follows.
\begin{itemize}
\item The set of provided service qualities is $\mathcal{Q}=q([0,1])$.
\item The pricing strategy is $p(q)=\vartheta v(q)-\int\nolimits_{0}^{\vartheta} v(q(s))\mathrm{d}s-V(0)$, where $\vartheta$ is the type that satisfies $q(\vartheta)=q$.
%\item The capacity allocation strategy is $c(q)=w(q)\int_{\Phi(q)}f(\theta)\mathrm{d}\theta $ , where $\Phi(q) \triangleq \{\theta \in \Theta : q(\theta)=q\}$.
\end{itemize}
Given the above discussions, the ISP's profit maximization problem boils down to finding the corresponding user's choice function $q^*(\cdot)$ and the indirect utility level $V^*(0)$ under the optimal service differentiation scheme $\mathcal{Q}^*$ and $p^*(\cdot)$\footnote{We take a detour to show the existence and uniqueness of optimal strategy. In the rest of the section, we first transform the ISP's profit-maximizing problem to an equivalent optimal control problem, and find out the unique candidate solution that satisfies the necessary condition, i.e., the  Pontryagin's Maximum Principle \cite{chiang2000elements}. Finally we use the sufficient conditions to prove that the solution is indeed the optimal strategy. We leave the details to the proof of Theorem \ref{fixedc_sol} in the Appendix.}.

\subsection{Optimal Control Framework}
We demonstrate that the ISP's profit maximization problem can be transformed into an optimal control problem, the solution to which is the user's choice function $q^*(\cdot)$ and the indirect utility level $V^*(\cdot)$ under the optimal service differentiation.

\begin{lemma}\label{v0}
Under the ISP's optimal service differentiation scheme, the initial level of indirect utility satisfies $V^*(0)=0$.
\end{lemma}

Lemma \ref{v0} states that under the ISP's optimal service differentiation, users with the lowest valuation, i.e., $\theta=0$, obtains zero utility. Consequently, we can set the indirect utility level $V(0)$ to be zero when searching for the optimal strategy.

To calculate the total profits of the ISP using $V(\cdot)$ and $q(\cdot)$, we can perform integration with respect to the user type $\theta$ instead of over the service class domain $\mathcal{Q}$, i.e.,
\begin{align*}
\int_{\mathcal{Q}} p(q)\left(\int_{\Phi(q)}f(\theta)\mathrm{d}\theta\right) \mathrm{d}q=\int_{0}^{1} [\theta v(q(\theta))-V(\theta)]f(\theta)\mathrm{d}\theta.\nonumber
\end{align*}
The consumed capacity can be calculated by integrating the unit capacity function $w(q)$ over the user types $\theta$ as follows.
\begin{lemma}\label{capacity}
The total  consumption of network capacity in the system can be calculated by
$\int\nolimits_{0}^{1}w(q(\theta))f(\theta)\mathrm{d}\theta$.
\end{lemma}

By Lemma \ref{capacity}, we can define a state variable
\begin{equation}\label{W_def}
W(\theta) \triangleq \int\nolimits_{0}^{\theta} w(q(s))f(s)\mathrm{d}s
\end{equation}
to measure the aggregate capacity consumption by users with valuation lower than $\theta$. In particular, $W(1)$ is the total capacity consumption of the entire system.

\begin{lemma}\label{fixed_control}
The user's choice function $q^*(\cdot)$ under the optimal service differentiation can be determined by solving the following optimal control problems. \\
\textit{a)} For the fixed capacity scenario:
\begin{align}\label{fixed_def}
\underset{q(\cdot)}{\text{maximize}}\quad &J=\int_{0}^{\theta_{M}} [\theta v(q(\theta))-V(\theta)]f(\theta)\mathrm{d}\theta \nonumber\\
\text{subject to}\quad
&q(\cdot)\ \ \text{nonincreasing}, \ \  q(\theta) \in [0,1] \ \forall \theta, \nonumber \\
&V'(\theta)=v(q(\theta)), \quad V(0)=0, \nonumber\\
&W'(\theta)=w(q(\theta))f(\theta), \quad W(0)=0, \nonumber\\
& \text{and} \ W(\theta_{M})\leq C_{M}.
\end{align}
b) For the variable capacity scenario:
\begin{align}\label{expansion_def}
\underset{q(\cdot)}{\text{maximize}}\quad &J=\int_{0}^{1} [\theta v(q(\theta))-V(\theta)]f(\theta)\mathrm{d}\theta
-S(W(\theta_{M})) \nonumber\\
\text{subject to}\quad
&q(\cdot)\ \ \text{nonincreasing}, \ \  q(\theta) \in [0,1] \ \forall \theta, \nonumber \\
&V'(\theta)=v(q(\theta)),\quad V(0)=0, \nonumber\\
&W'(\theta)=w(q(\theta))f(\theta),\quad W(0)=0.
\end{align}
\end{lemma}

The above problems (\ref{fixed_def}) and (\ref{expansion_def}) are  more tractable than the original ones (\ref{fixed_cap}) and (\ref{capexpansion}). In these two optimal control problems, the user's choice function $q(\cdot)$ is the control variable in the functional space $\{q(\cdot) \in \mathbb{R}^{[0,1]} | q(\cdot) \ \text{nonincreasing}\}$ and the indirect utility function $V(\cdot)$ and capacity consumption function $W(\cdot)$ are the so-called state variables.
%We can decompose the dynamic optimization problems into a family of static optimization problems using the Pontryagin's Maximum Principle \cite{chiang2000elements}.

\subsection{Optimal Solution}
Notice that the monotonicity constraint on $q(\cdot)$ requires special treatments. In particular, we adopt the following steps.
\begin{itemize}
\item We first consider a relaxed optimal control problem in which the monotonicity constraint is removed. By using the Pontryagin's Maximum Principle \cite{chiang2000elements}, we find a candidate optimal control solution  $\widetilde{q}(\cdot)$.
%\footnote{There might be no such candidate or more than one. Here we assume that an unique $\widetilde{q}(\cdot)$ satisfies the Maximum Principle for simplicity of introducing our steps.}.
\item We check whether $\widetilde{q}(\cdot)$ satisfies the sufficient conditions for optimality. If so, $\widetilde{q}(\cdot)$ is indeed an optimal control of the relaxed problem and we proceed to the next step.
\item If $\widetilde{q}(\cdot)$ meets the monotonicity constraint, then it is also an optimal control of the original problem (\ref{fixed_control}), i.e. $q^*(\cdot)=\widetilde{q}(\cdot)$; otherwise, we conduct further analysis based on $\widetilde{q}(\cdot)$.
\end{itemize}

We use the fixed capacity scenario to illustrate the key steps in solving the relaxed optimal control problem. We first define the Hamiltonian $H$ as follows:
\begin{align}\label{H}
&H[\theta,q(\cdot),V(\cdot),W(\cdot),\lambda_1(\cdot),\lambda_2(\cdot)] \\
&\triangleq  [\theta v(q(\theta))\!-\!V(\theta)]f(\theta)\!+\!\lambda_1(\theta)v(q(\theta))\!+\!\lambda_2(\theta)w(q(\theta))f(\theta),\nonumber
\end{align}
where $\lambda_1(\cdot)$ and $\lambda_2(\cdot)$ are the so-called co-state variables satisfying the following transversality conditions:
\begin{align*}
\begin{cases}
\displaystyle\frac{d\lambda_1}{d\theta}=-\frac{\partial H}{\partial V}=f(\theta)\vspace{0.05in}\\
\displaystyle\frac{d\lambda_2}{d\theta}=-\frac{\partial H}{\partial W}=0\\
\displaystyle\lambda_1(1)=0, \quad \lambda_2(1)\leq 0, \quad \lambda_2(1)[W(1)-C_M]=0.
\end{cases}
\end{align*}
It immediately follows that $\lambda_1(\theta)=F(\theta)-1$ and $\lambda_2(\theta)$ is a constant over $[0,1]$. If all the network capacity is consumed, i.e., $W(1)-C_M=0$, the slackness condition gives $\lambda_2(1)<0$; otherwise $\lambda_2(1)=0$. In either case, we define $\lambda_2(\theta)=-\mu$ where $\mu$ is a nonnegative real constant.

The economic interpretation of the Hamiltonian and the co-state variables are as follows. The co-state constant $\mu$ is the \textit{shadow price} of the network capacity which evaluates the contribution to total profits by consuming per unit capacity. The Hamiltonian $H$ is the \textit{surrogate profit} the ISP extracts from the users of type $\theta$, which consists of three components on the right-hand side of (\ref{H}). The first component is the direct impact of the control variable $q(\cdot)$ on the object function $[\theta v(q(\theta))-V(\theta)]f(\theta)$. The control variable $q(\theta)$ also has an indirect impact on the object function: it influences the value of $V(\theta+\delta\theta)$ in the next infinitesimal type by the differential $V'(\theta)=v(q(\theta))$, which is captured by the second component. Moreover, the third component characterizes the cost of capacity consumption.

According to the Pontryagin's Maximum Principle, if an optimal control $\widetilde q(\cdot)$ exists, then along the optimal trajectories, $\widetilde q(\cdot)$ is the point-wise maxima of the Hamiltonian $H$:
\begin{equation}\label{newH}
[\theta v(q)-\widetilde V(\theta)]f(\theta)+[F(\theta)-1]\theta v(q)-\mu  w(q)f(\theta),
\end{equation}
where $\widetilde V(\cdot)$ is the optimal trajectory of state. The economic interpretation is that to maximize its total profits, the ISP has to maximize the surrogate profit it extracts from each type of users. Neglecting the components irrelevant to $q$ in (\ref{newH}), we see that $\widetilde q(\cdot)$ is the point-wise maxima of \(\Psi(\theta,q)\), defined by
\begin{equation}\label{psi}
\Psi(\theta,q)\triangleq\left(\theta-\frac{1-F(\theta)}{f(\theta)}\right)v(q)-\mu  w(q).
\end{equation}

By far we have illustrated the first step, i.e., to use the Maximum Principle to find the possible candidate optimal control $\widetilde q(\cdot)$. Due to space limitation, we show the remaining steps in the proof of Theorem \ref{fixedc_sol} and proceed to present our results, i.e., $q^*(\cdot)$ and the corresponding optimal service differentiation $\mathcal{Q}^*$ and $p^*(\cdot)$. We introduce the following Assumption \ref{G} and \ref{h} to facilitate further analysis.

\begin{assumption}\label{G}
The \textit{virtual valuation} function $G(\theta) \triangleq \theta-\frac{1-F(\theta)}{f(\theta)}$, which is the coefficient of $v(q)$ in (\ref{psi}), has a unique zero point $\theta_{0}$ on $(0,1)$ and is strictly increasing on $(\theta_{0},1)$.
\end{assumption}

The virtual valuation function $G(\cdot)$ is first defined by Myerson in \cite{myerson1981optimal} designing revenue-optimal auctions. One sufficient condition for Assumption \ref{G} is the standard \textit{monotone hazard rate} assumption in the mechanism design literature such as \cite{hartline2009simple}. Assumption \ref{G} is satisfied by many common distributions, e.g., normal, exponential and beta distributions. We denote $G^{-1}(\theta)$ as the inverse function of $G(\theta)$ over the interval $(\theta_{0},1)$.

\begin{assumption}\label{h}
We define $h(q)\triangleq w'(q)/v'(q)$ as a \textit{virtual capacity} function  and assume that it is decreasing in \(q\) over the interval $(0,1)$.
\end{assumption}

The virtual capacity function $h(\cdot)$ is defined as the ratio of the marginal unit capacity $w'(q)$ needed to support service quality at a congestion level $q$ to the marginal value of a user $\theta$ whose best choice of quality is $q$.
%has a clear economic meaning.
%For any given type of user $\theta$, the ratio $\frac{w'(q)}{\theta v'(q)}=\frac{h(q)}{\theta}$ is the extra capacity required of service class with congestion level $q$ to improve the user's achieved valuation by one unit.
$h(q)$ measures the marginal capacity needed so as to marginally increase the user value at the level $q$ of congestion in the system.
Assumption \ref{h} intuitively states that more capacity is required of the service class with milder congestion to further improve users' achieved values. Furthermore, we define a function $h^{-1}(\cdot)$ over the interval $[h(1),+\infty)$, whose value is the inverse of $h(\cdot)$ over $[h(1),h(0)]$ and zero over $(h(0),+\infty)$.

Under Assumption \ref{G} and \ref{h}, we are able to derive a concise analytic solution of the optimal control problem and simplify the presentation of our results as follows. %Under the assumptions, the optimal control $\widetilde q(\cdot)$  obtained from (\ref{psi}) is monotonic and we have $q^*(\cdot)=\widetilde q(\cdot)$.
%We defer the discussions when Assumption \ref{G} and \ref{h} are relaxed to the Appendix.

\begin{theorem}\label{fixedc_sol}
For the fixed capacity scenario, let $\bar C \triangleq [1-F(\theta_0)]w(0)$.
If $C_M < \bar C$, the user choice function $q^*(\cdot)$ and the shadow price $\mu$ are uniquely determined by the following set of equations.\\
\textit{a)} The user choice function:
\[ q^*(\theta)=\begin{cases}
      1, & \forall \theta \in [0,\hat\theta); \\
      h^{-1}(G(\theta)/\mu), & \forall \theta \in [\hat\theta,1]. \\
   \end{cases}
\]
\textit{b)} The \textit{marginal user}\footnote{The marginal user $\hat\theta$ is the type of user that is indifferent between joining and opting-out of the services. Users with valuations below $\hat\theta$ do not join the services while users with valuations above $\hat\theta$ do.} $\hat\theta$ that uses the network services:
\begin{displaymath}
\hat\theta=G^{-1}(\mu \cdot h(1)).
\end{displaymath}
\textit{c)} The capacity equation:
\begin{displaymath}
\int\nolimits_{0}^{1}w(q^*(\theta))f(\theta)\mathrm{d}\theta = C_{M}.
\end{displaymath}
Otherwise, the ISP's optimal differentiation is to provide a single-class service, which satisfies \(\mathcal{Q}=\{0\}\) and \(p^*(0)=\theta_0\).
\end{theorem}

The upper bound $\bar C$ defined in Theorem \ref{fixedc_sol} is a threshold that determines whether the system's capacity is \textit{abundance}. Notice that the parameterization of our model is quite flexible. When $w(0)=+\infty$, we have $\bar{C}=+\infty$. In this case, $q=0$ is unreachable, which models the situation where the user always perceive the congestion externality. On the other hand, when $w(0)<+\infty$, we have $\bar{C} < +\infty$. This setting can be used to model the situation where users perceive no congestion externality when congestion level drops below certain threshold. From Theorem \ref{fixedc_sol}, we see that if the ISP's network capacity is limited, i.e. $C_M < \bar C$, all the inequalities in (\ref{cap_constraint}) will be  binding under the ISP's optimal service differentiation.

%\begin{align}
%\begin{split}
%&c_{i}=\hat c_{i}=C(d_{i}(\mathbf{p},\mathbf{q}),q_{i}), \quad i\in\mathcal{N}\\
%&\textstyle\sum_{i \in \mathcal{N}} C\big(d_{i}(\mathbf{p},\mathbf{q}),q_{i}\big)=C_M
%\end{split}
%\end{align}

Theorem \ref{fixedc_sol} characterizes the optimal user choice function $q^*(\cdot)$ for the fixed capacity scenario. With necessary modifications, we can apply similar procedures to derive the solution for the variable capacity scenario as follows.

\begin{theorem}\label{expansion}
For the variable capacity scenario, if the cost function $S(\cdot)$ is convex with $S(0)=0$, the optimal user choice function $q^*(\cdot)$ and the associated capacity consumption $W^*(1)$ are uniquely determined by the following set of equations.\\
\textit{a)} The user choice function:
\[ q^*(\theta)=\begin{cases}
      1, & \forall \theta \in [0,\hat\theta); \\
      h^{-1}(G(\theta)/S'(W^*(1))), & \forall \theta \in [\hat\theta,1]. \\
   \end{cases}
\]
\textit{b)} The \textit{marginal user} $\hat\theta$ that uses the network services:
\begin{displaymath}
\hat\theta=G^{-1}\left(S'\big(W^*(1)\big)  h(1)\right).
\end{displaymath}
\textit{c)} The capacity equation:
\begin{displaymath}
\int\nolimits_{0}^{1}w(q^*(\theta))f(\theta)\mathrm{d}\theta = W^*(1).
\end{displaymath}
\end{theorem}
After we have obtained $q^*(\cdot)$ by the above theorems, the optimal service differentiation scheme becomes straightforward.
\begin{corollary}\label{optimalstrategy}
Under the ISP's optimal service differentiation, the service classes form an interval $\mathcal{Q^*}=[q^*(1),q^*(\hat\theta)]$, and the associated prices satisfy
\begin{displaymath}p^*(q)=\vartheta v(q)-\int\nolimits_{0}^{\vartheta} v(q^*(s))\mathrm{d}s, \quad \forall q \in \mathcal{Q},
\end{displaymath}
where $\vartheta$ is the user type such that $q^*(\vartheta)=q$. Furthermore, $p^*(\cdot)$ is continuously differentiable and decreasing in $q$.
\end{corollary}

\subsection{Characteristics of the Optimal Solution}
In this subsection, we study the structures and properties of the optimal service differentiation scheme and the corresponding user choices. We first concentrate on the fixed capacity scenario and the following corollary characterizes the shadow price $\mu$ of network capacity.
\begin{corollary}\label{shadow}
For any given user distribution $F(\cdot)$ satisfying Assumption \ref{G}, the shadow price $\mu$ decreases in the system capacity $C_M$ with the asymptotic limit $\lim_{C_M \to \bar C} \mu=0$.
\end{corollary}

Corollary \ref{shadow} intuitively states that the marginal value of capacity decreases when the ISP has more capacity, a common diminishing return effect, and this  marginal value drops to zero when the capacity becomes abundant.

This next result explicitly shows the type of users that will be served under an ISP's optimal service differentiation.

\begin{corollary}\label{exclude}
Under an ISP's optimal service differentiation, users with values lower than the marginal user value $\hat\theta$ will not use any network service. In particular, the marginal user value $\hat\theta$ satisfies that
\begin{itemize}
\item[\textit{a)}] if $h(1)=0$, $\hat\theta$ is the unique solution to the equation $G(\hat\theta)=\hat\theta-\frac{1-F(\hat\theta)}{f(\hat\theta)}=0$, i.e. $\hat\theta=\theta_0$.
\item[\textit{b)}] if $h(1)>0$, $\hat\theta$ depends on both the user distribution $F(\cdot)$ and the fixed capacity $C_M$. Given any $F(\cdot)$, $\hat\theta$ decreases in $C_M$ with the asymptotic limit $\lim_{C_M \to \bar C}=\theta_0$.
\end{itemize}
\end{corollary}
Corollary \ref{exclude} states that the optimal service differentiation offers the service classes with sufficiently low levels of congestion and high prices such that the users with low value will not use any service, even when the network capacity is abundant. %In particular, we have the following corollary.
The intuition behind Corollary \ref{exclude} is that by offering premium services with high quality and high prices, the gains from the high-end users outweigh the loss of low-end market share.

Next, we turn to the structure of the optimal choices of users with values $\theta$ above $\hat\theta$. Letting $\bar\theta \triangleq \min\{G^{-1}(\mu  h(1)),1\}$, we rewrite their choices $q^*(\theta)$ as follows.
\[ q^*(\theta) = \begin{cases}
      h^{-1}\left(\displaystyle\frac{G(\theta)}{\mu}\right), & \forall \theta \in [\hat\theta,\bar\theta]; \\
      0, & \forall \theta \in [\bar\theta, 1].
   \end{cases}
\]
Over the interval $[\hat\theta,\bar\theta]$, a user of type $\theta$ joins the service class $q^*(\theta)$ such that $G(\theta)=\mu  h(q^*(\theta))$, i.e., the user's virtual valuation $G(\theta)$ equals the cost of maintaining service class $q^*(\theta)$, which is the shadow price multiplied by the virtual capacity. In other words, under the ISP's optimal service differentiation, each user of type $\theta$ will choose the service class with highest quality and price that it can afford.
Moreover, when $h(0)<+\infty$, the cost of maintaining services is upper bounded by $\mu h(0)$, and the user's virtual valuation might be higher than this bound. In this case, high-end users will be bunched up in the service class $q=0$. The following corollary characterizes this bunching phenomenon.

\begin{corollary}\label{bunching}
When $h(0)<+\infty$, there exists a threshold $\hat C < \bar C \triangleq [1-F(\theta_0)]w(0)$ such that
\begin{itemize}
\item [\textit{a)}] If $C_M \leq \hat C$,  $q(\theta)$ is strictly decreasing in $[\hat\theta,1]$, implying that  all the users that actually choose the ISP's services are completely differentiated under the optimal strategy.
\item [\textit{b)}] If $\hat C < C_M <\bar C$, we have $\bar\theta<1$ and $q(\theta)=0$ for all $\theta \in [\bar\theta, 1]$, implying that high-end users are bunched up in the best service class with $q=0$.
\end{itemize}
When $h(0)=+\infty$, $q(\theta)$ is strictly decreasing on $[\hat\theta,1]$, implying again that users are completely differentiated.
\end{corollary}

We can analyze the solution for the variable capacity scenario in a similar way. In Theorem \ref{expansion}, the marginal price of network capacity at the optimal consumption level $W^*(1)$ is $S'(W^*(1))$, which plays the same role as the shadow price $\mu$ in Theorem \ref{fixedc_sol} and depends on both the user distribution $F(\cdot)$ and the cost function $S(\cdot)$. Therefore, the above discussions can be readily extended to handle the variable capacity scenario.

In summary, Corollaries \ref{exclude} and \ref{bunching} show that
%$h(\cdot)$ plays an important role in the structures of the optimal strategy.
%Since the network capacity in a system with $h(1)>0$ or $h(0)=+\infty$ is in general more expensive, we can summarize the implications as follows.
an ISP's optimal service differentiation scheme always offers service classes with sufficiently low levels of congestion and high prices such that users with low values do not subscribe to the ISP, even if it has abundant network capacity. If the ISP expands its capacity, its market share should be increased; if the ISP's network capacity is sufficient, i.e., more than $\hat C$, it is optimal to bunch up the high-end users in the best service class with $q=0$. Furthermore, when the users' valuations lean towards high values, the ISP needs to sacrifice part of the low-end market share and dedicate its capacity to premium service classes for high-end users so as to maximize its profits.

\section{Dynamics of Optimal Service Differentiation}
In this section, we study the dynamics of an ISP's optimal service differentiation and the corresponding user behavior. We first choose the parameters of the system. In particular, we use the exponential form $e^{-q}$ to construct the user's satisfaction discount on congestion level. As we normalize the domain of congestion $q$ to be $[0,1]$, we adopt a normalized form $v(q)=\frac{e^{-q}-e^{-1}}{1-e^{-1}}$ which satisfies  $v(0)=1$ and $v(1)=0$. We further model the implied capacity by the quadratic convex function $w(q)=(1-q)^2$, under which the marginal demand of capacity is increasing as the required service quality becomes better, i.e., as $q$ decreases. Under these settings, the virtual capacity function satisfies $h(0)<+\infty$ and $h(1)=0$.

\subsection{Impacts of Network Capacity Constraint}
We study the dynamics of the ISP's optimal service differentiation with respect to network capacity constraint, i.e., the maximum capacity $C_{M}$ in the fixed capacity scenario and the cost function $S(\cdot)$ in the variable capacity scenario.
\begin{figure}[ht]
	\includegraphics[width=0.240\textwidth]{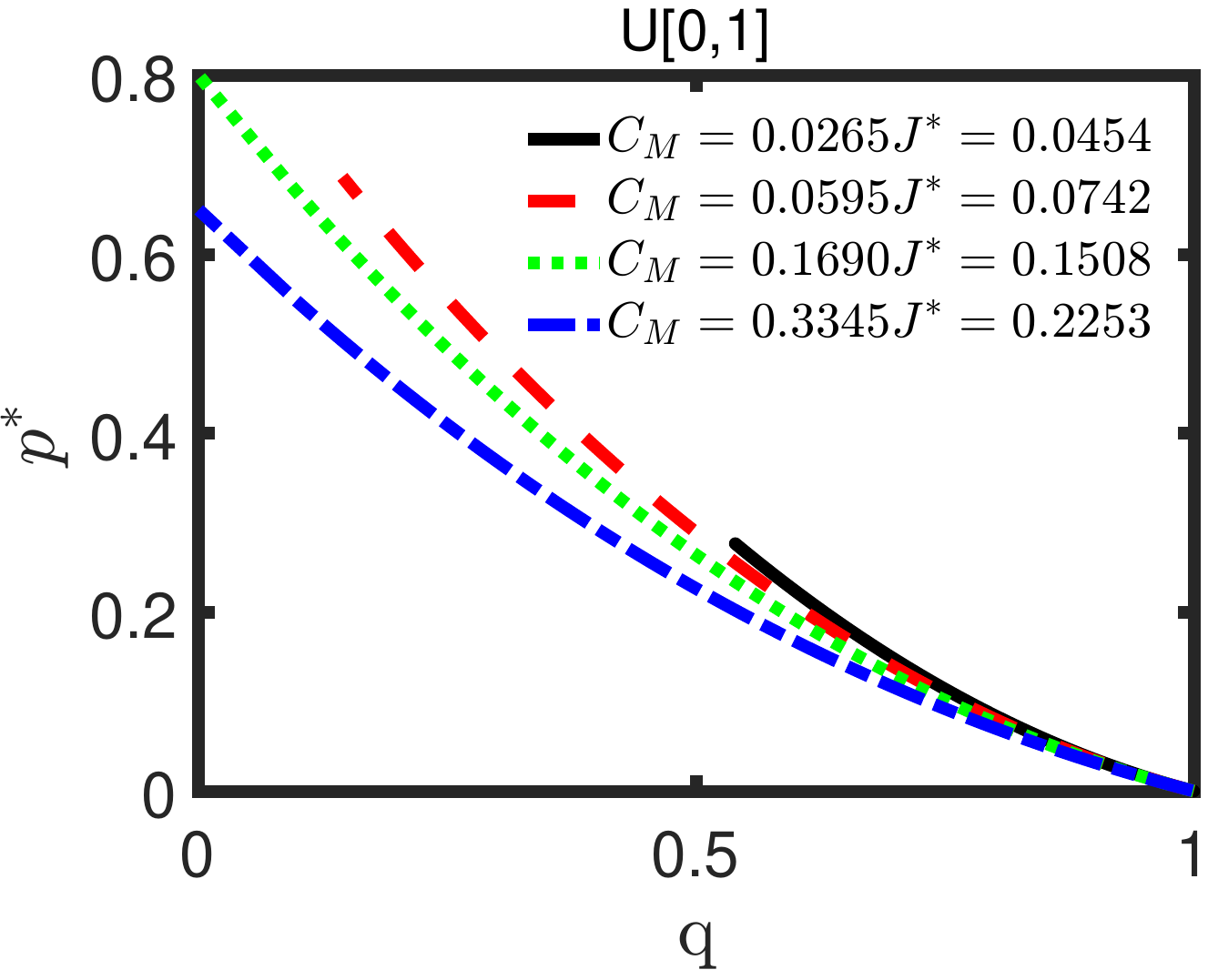}
	\includegraphics[width=0.243\textwidth]{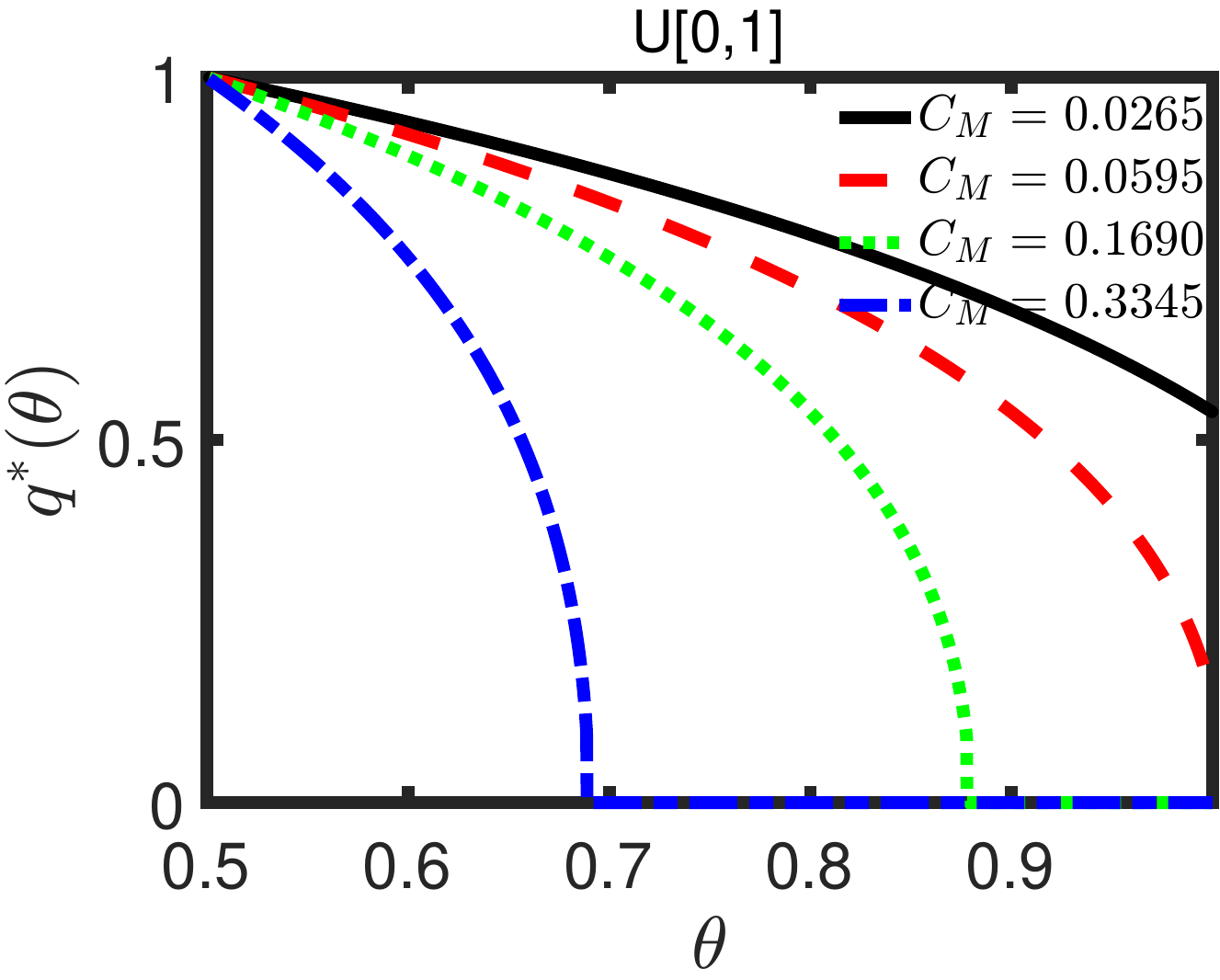}
	%\vspace{-2em}
	\caption{Optimal pricing under different maximum capacity}
	\vspace{-1em}
	\label{impact_C}
\end{figure}

We first investigate the former one. Fig \ref{impact_C} plots the ISP's optimal service differentiation (on the left) and the corresponding user's choice function (on the right) when user type $\theta$ follows the uniform distribution $U[0,1]$. Each curve in the figures represents a different $C_M$ and $J^*$ is the corresponding optimal profits. We observe that when the maximum capacity $C_{M}$ increases, 1) both the ISP's optimal pricing curve and the user choice curve shift downwards, 2) only users with valuation $\theta \in [0.5,1]$ choose the services, 3) when capacity is sufficient, high-end users are bunched onto the service class $q=0$, and 4) the ISP's profit rises. The second observation is a demonstration of Corollary \ref{exclude} that when $h(1)=0$, the marginal user of type $\hat\theta$ is independent of $C_M$.
Besides, the observed bunching phenomenon is an illustration of Corollary \ref{bunching}. We summarize the other observations in the following corollaries.

\begin{corollary}\label{capacity_theorem}
Under a fixed user distribution $F(\cdot)$, denote the ISP's optimal range of service qualities under the maximum capacity $C_{M1}$ and $C_{M2}$ as $\mathcal{Q}_{1}$ and $\mathcal{Q}_{2}$, respectively. If $C_{M1}<C_{M2}$, then $\mathcal{Q}_{1} \subset \mathcal{Q}_{2}$.\footnote{We also observe that the optimal prices satisfy $p_{2}(q)<p_{1}(q)$ for all $q \in \mathcal{Q}_{1} \cap \mathcal{Q}_{2}\backslash\{1\}$. However, this observation has not been rigorously proved.}
\end{corollary}

Corollary \ref{capacity_theorem} indicates that when the ISP's maximum capacity is larger, its optimal service differentiation offers more premium classes. We also observe that the ISP should decrease the prices to maximize its profit under this scenario.

\begin{corollary}\label{profits}
Under a fixed user distribution $F(\cdot)$, an ISP's optimal profit $J^*$ increases with its maximum network capacity $C_{M}$ over $(0,\bar C)$ and have the following asymptotic limit:
\begin{equation}
\lim_{C_{M}\to \bar C} J^*=\theta_{0}[1-F(\theta_{0})].
\end{equation}
\end{corollary}

Corollary \ref{profits} states that if an ISP's capacity is not abundant, i.e., $C_M<\bar C$, it can gain more profits under the optimal service differentiation when its maximum capacity expands.

Next we study the variable capacity scenario. We focus on the impacts of cost function $S(\cdot)$ on the optimal consumption $W^*(1)$ of capacity and the profits $J$. To this end, we choose a family of cost functions parameterized by a single parameter $t$ : $S(c)\triangleq 0, \forall c \in [0,0.1]$ and $S(c) \triangleq t(c-0.1)^2, \forall c \in (0.1,+\infty)$. Our settings model the scenario where the ISP's existing capacity equals $C=0.1$ and only need to pay for costs of the additional capacity. In general, a larger $t$ measures higher costs for expanding capacity.

\begin{figure}[t]
\includegraphics[width=0.24\textwidth]{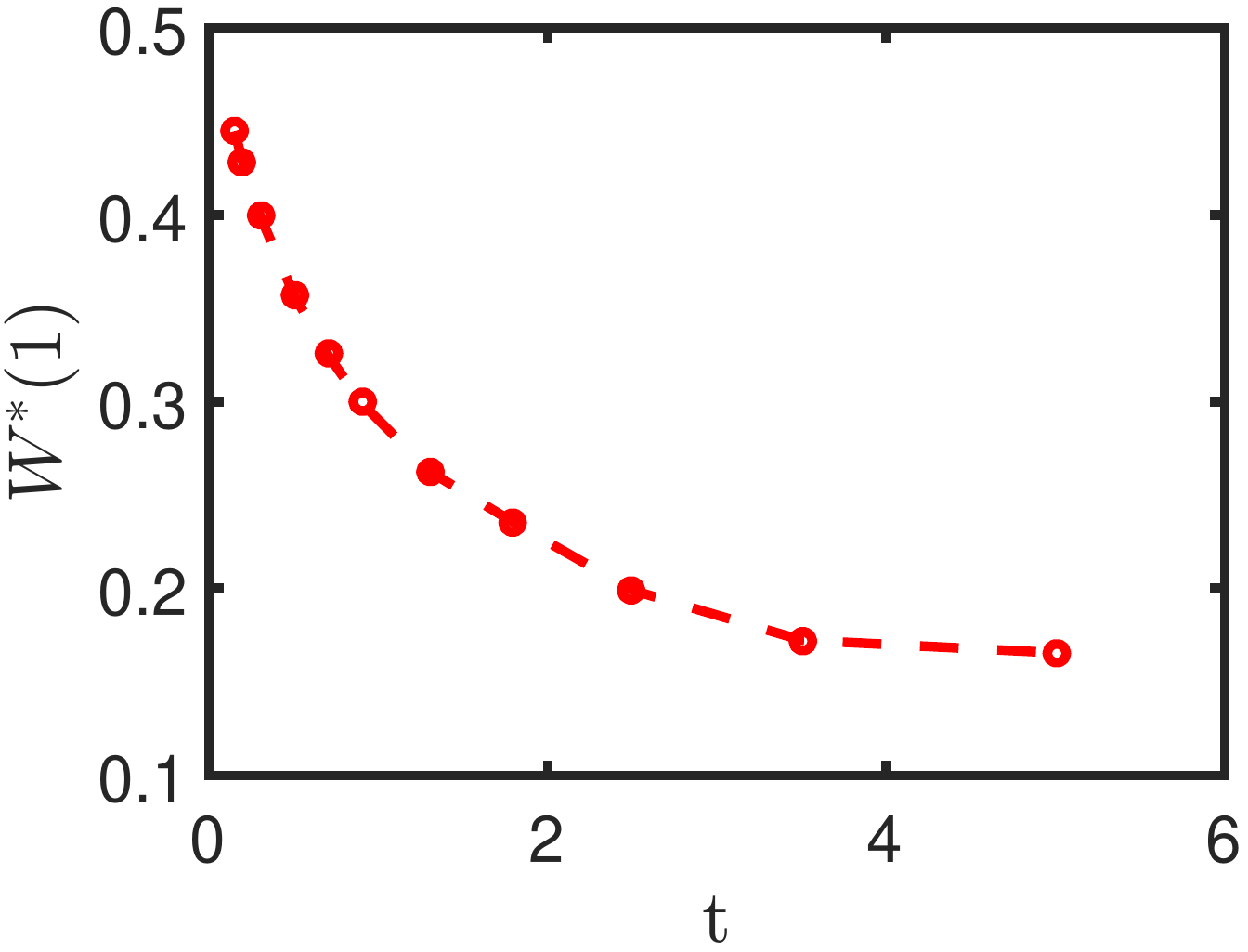}
\includegraphics[width=0.24\textwidth]{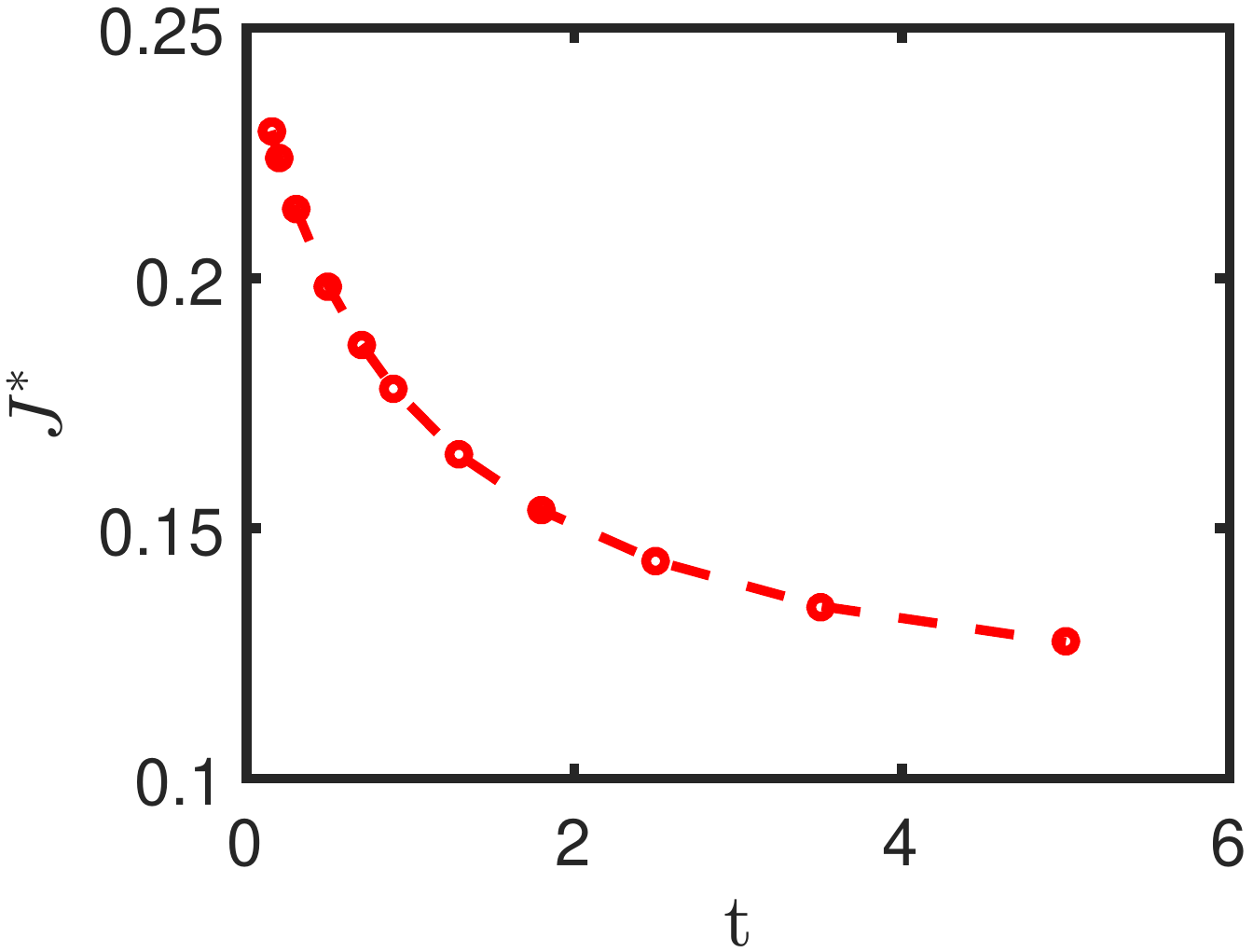}
\vspace{-2em}
\caption{Impacts of the cost function}
\vspace{-1em}
\label{impact_S}
\end{figure}

Figure \ref{impact_S} plots the optimal consumption $W^*(1)$ of capacity (on the left) and the optimal profits $J^*$ (on the right) versus the parameter $t$ varying along the x-axis. We observe that both $W^*(1)$ and $J^*$ decrease with $t$, which indicates that when the cost of expanding capacity is cheaper, it is optimal for the ISP to purchase more capacity so as to gain more profits. In particular, when $t \rightarrow +\infty$, the ISP will not invest in any extra capacity and the optimal consumption level is $W^*(1)=0.1$.

\textbf{Summary of implications:} In the fixed capacity scenario, when the ISP increases network capacity, its optimal service differentiation scheme will \textit{a)} enlarge the range of service congestions by offering more premium services, and \textit{b)} charge each service class a lower price. In this scenario, the ISP is able to allocate more capacity for the premium services and attract high-end users to use them so as to increase profits. In the variable capacity scenario, when the cost of capacity is cheaper, the ISP has incentives to purchase more capacity to obtain higher profits.

\subsection{Impacts of User Distribution}
In this subsection, we study the dynamics of the ISP's optimal service differentiation with respect to the user distribution. In particular, we consider a family of distribution functions $F(\theta)=\theta^\alpha$ for $\theta \in [0,1]$. If $\alpha=1$, the users' values follow the uniform distribution $U[0,1]$; otherwise, the users are either leaning towards higher  ($\alpha>1)$ or  lower  ($\alpha<1$) values.

\begin{figure}[t]
\includegraphics[width=0.238\textwidth]{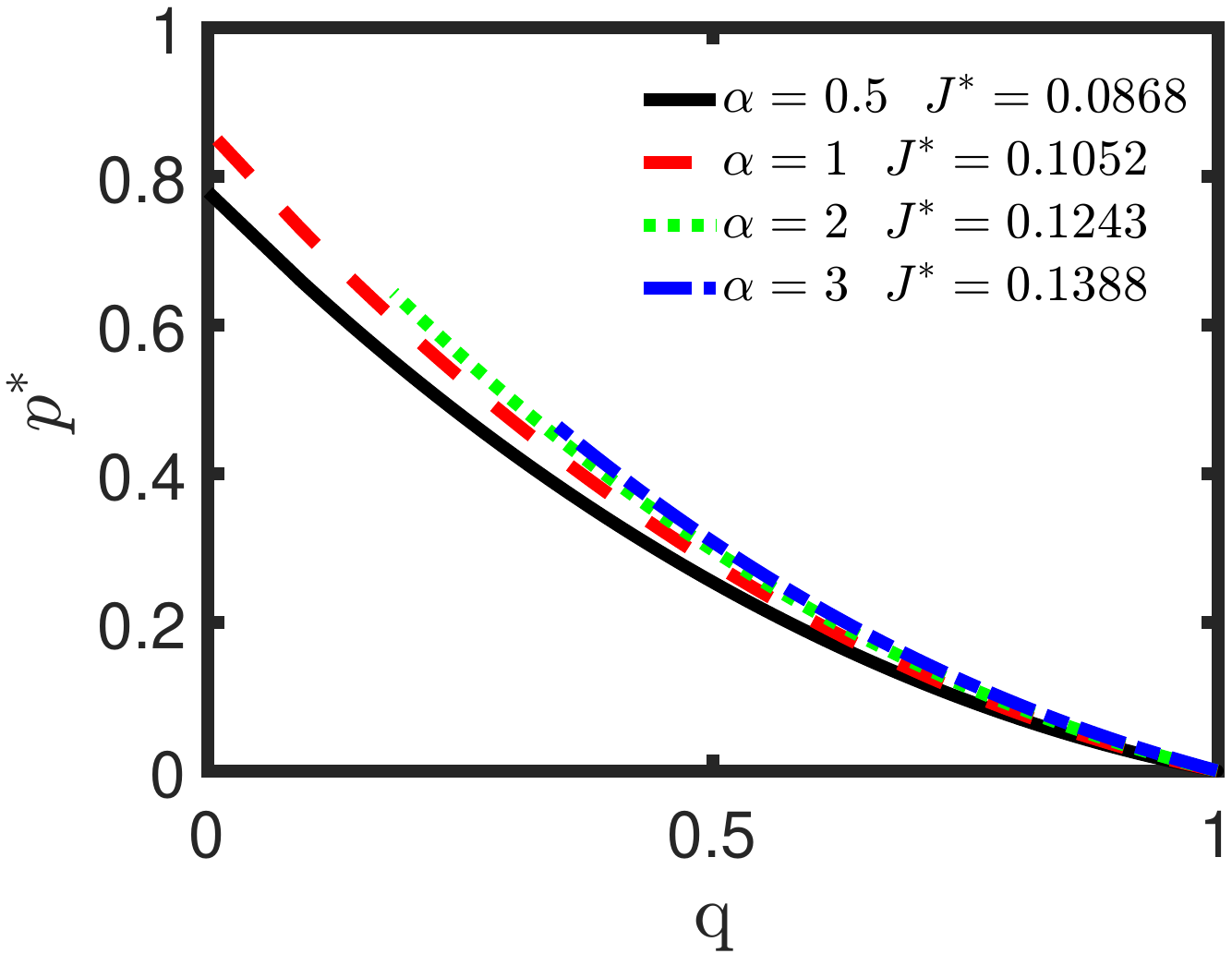}
\includegraphics[width=0.24\textwidth]{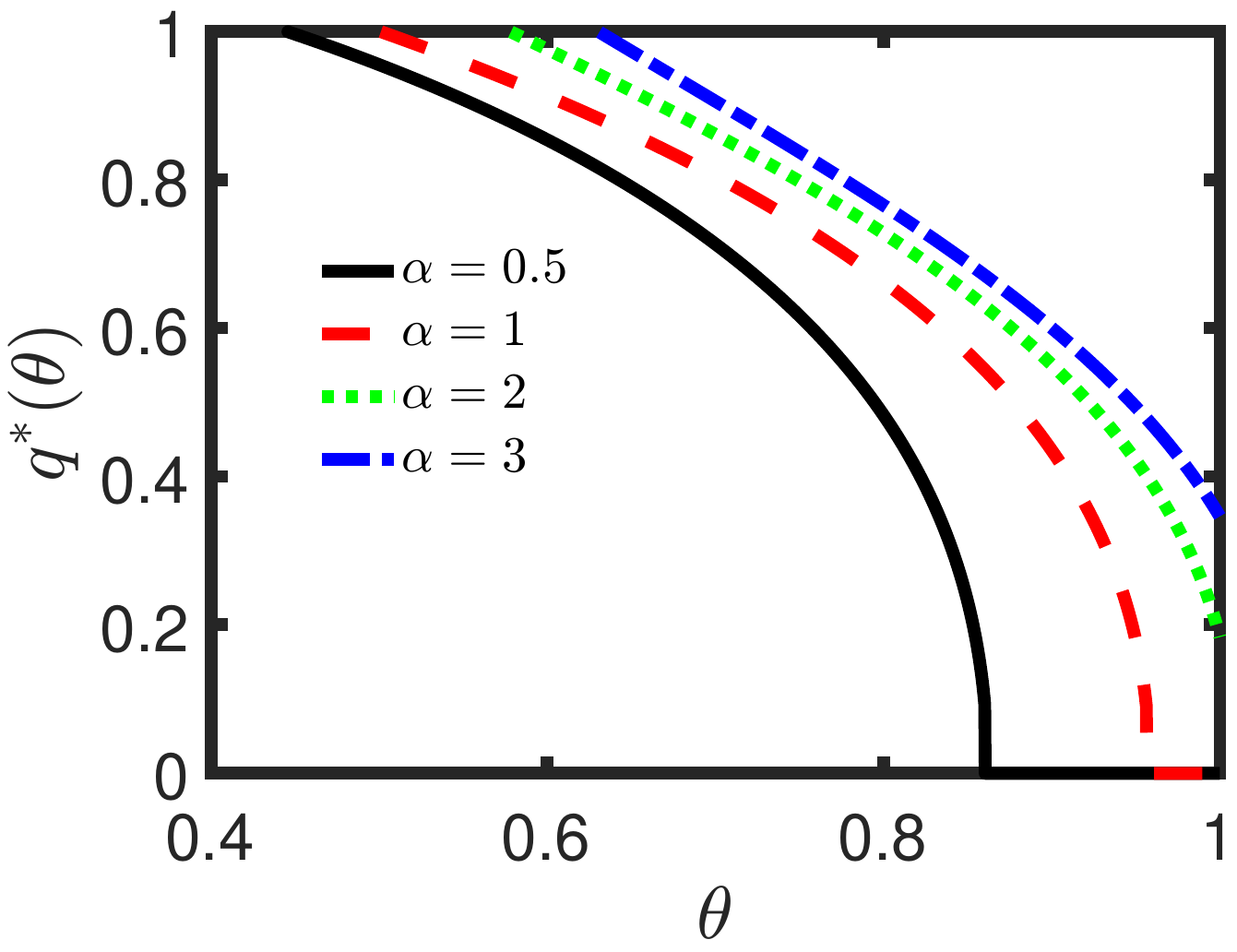}
\vspace{-0.5em}
\caption{Optimal pricing under different user distributions}
\vspace{-1em}
\label{dist_compare}
\end{figure}

Fig \ref{dist_compare} plots the ISP's optimal pricing curves (left) and the corresponding user choice curves (right) under various user distributions and fixed maximum capacity $C_{M}=0.1$. We observe that when user valuation shifts towards the higher end, 1) the ISP's optimal pricing curve and the corresponding user choice curve both shift upwards, 2) the ISP offers a narrower range of low congestion services and fewer users choose the network service,  3) the bunching effect vanishes when $\alpha$ is large, and 4) the ISP's profit rises. The second observation is a demonstration of Corollary \ref{exclude}, i.e., when $\alpha$ increases, the marginal user value $\hat\theta$ increases. The third observation is due to the fact that the threshold $\bar C\triangleq [1-F(\theta_0)]w(0)$ increases with the parameter $\alpha$.

\textbf{Summary of implications:} When there are more high-value users in the market, the ISP's optimal service differentiation will focus on these high-value users by offering a narrower range of premium service classes and charging higher prices, thus extracting more profits from these high-value users.

\subsection{Comparison with the Optimal Single-Class Service}
We evaluate whether an ISP has the incentive to implement service differentiation so as to obtain higher profits. To this end, we use the optimal  pricing strategy of a single-class service\footnote{The optimal single-class pricing strategy $(p^*,q^*)$ can be solved by restricting $|\mathcal{N}|=1$ in the optimization problem (\ref{fixed_cap}).}
$(p^*,q^*)$  as our benchmark and compare the ISP's profits and the total user surplus under our optimal service differentiation with that under the benchmark. We conduct simulations under the various user distributions. Since the general trends are almost the same, we adopt the uniform distribution $U[0,1]$ as the representative setting.

\begin{figure}[t]
\includegraphics[width=0.239\textwidth]{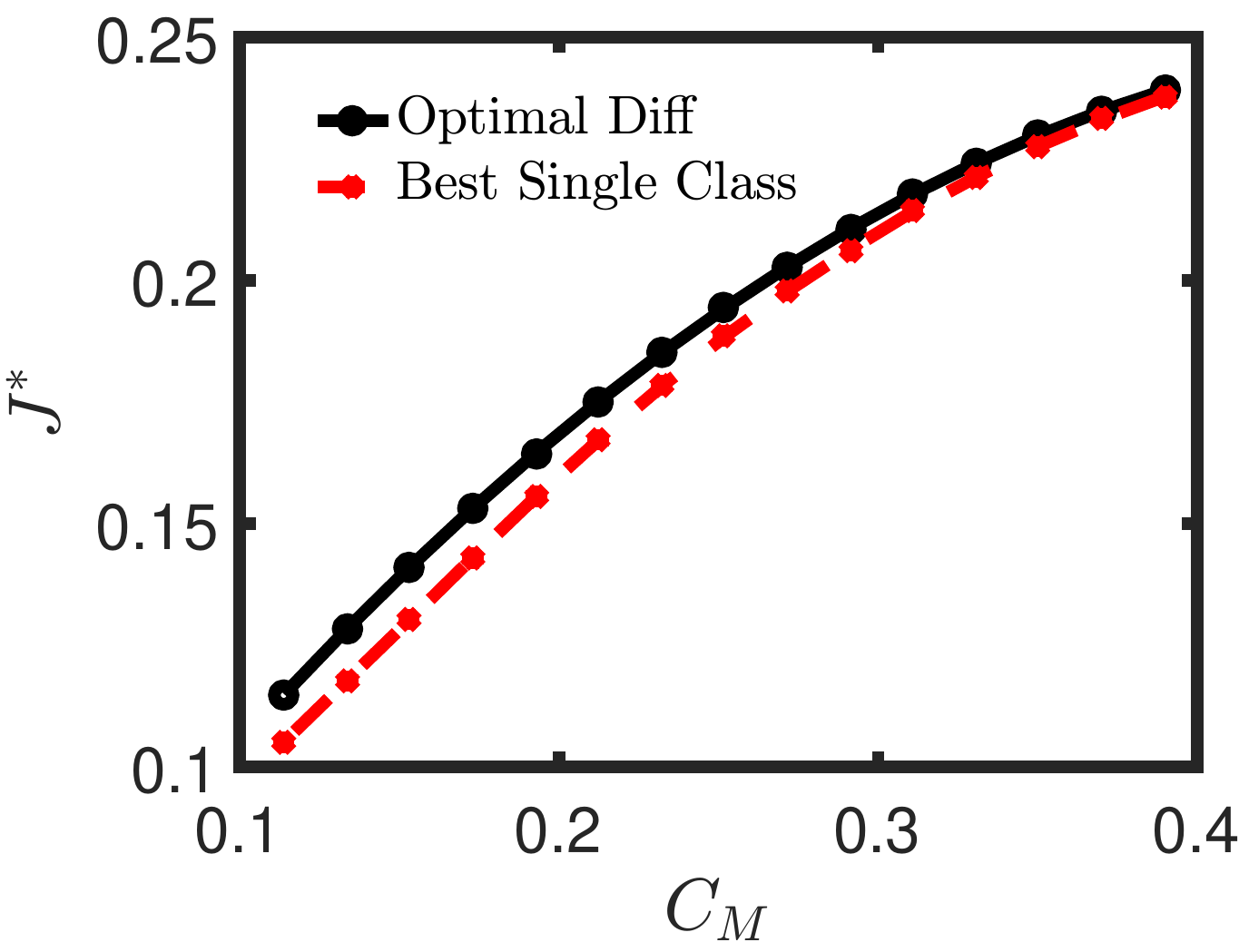}
\includegraphics[width=0.241\textwidth]{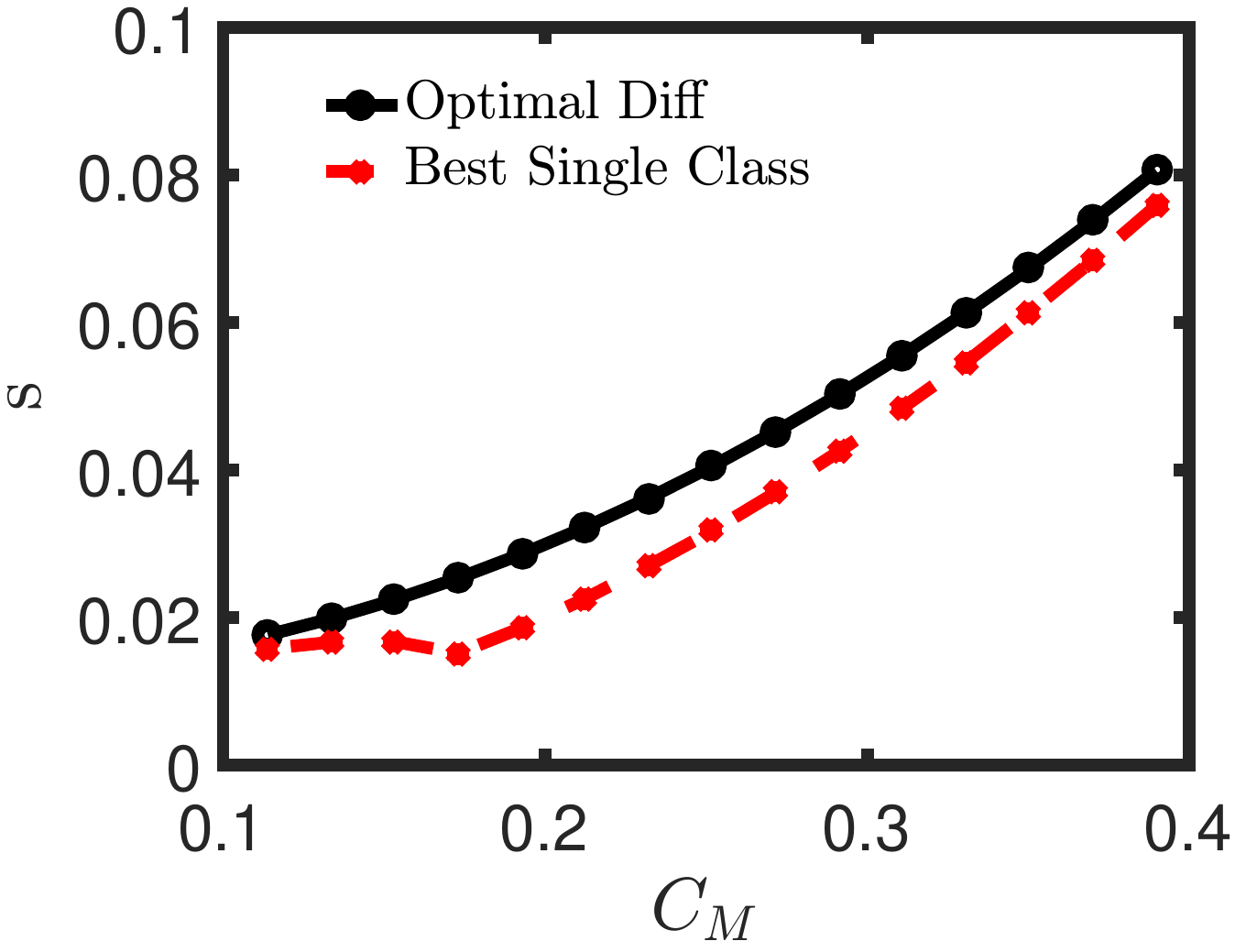}
\vspace{-2em}
\caption{Benefits of Service Differentiation}
\vspace{-1em}
\label{single_vs_opt}
\end{figure}

Figure \ref{single_vs_opt} plots the ISP's profits $J^*$ (on the left) and the total user surplus $s \triangleq \int_{0}^{1}V(\theta)\mathrm{d}\theta$ (on the right) versus the maximum capacity $C_M$ under the two strategies. We observe that when the capacity is scarce, the ISP has stronger incentives to differentiate services so as to gain more profits. However, as the capacity keeps growing, the increase in profits becomes marginal. This observation is consistent with Corollary \ref{profits} and Theorem \ref{fixedc_sol}, which state that when $C_M$ approaches $\bar C$, the ISP's profit converges to the upper bound which is reached without the capacity constraints. Figure \ref{impact_C} also illustrates this: the curve with $C_M=0.3345$ shows that when capacity expands, the majority of users are bunched up in the best service class $q=0$ and the profit gained by service classes $q \in (0,1]$ is small compared to that of the service class $q=0$.

\begin{figure}[t]
\includegraphics[width=0.24\textwidth]{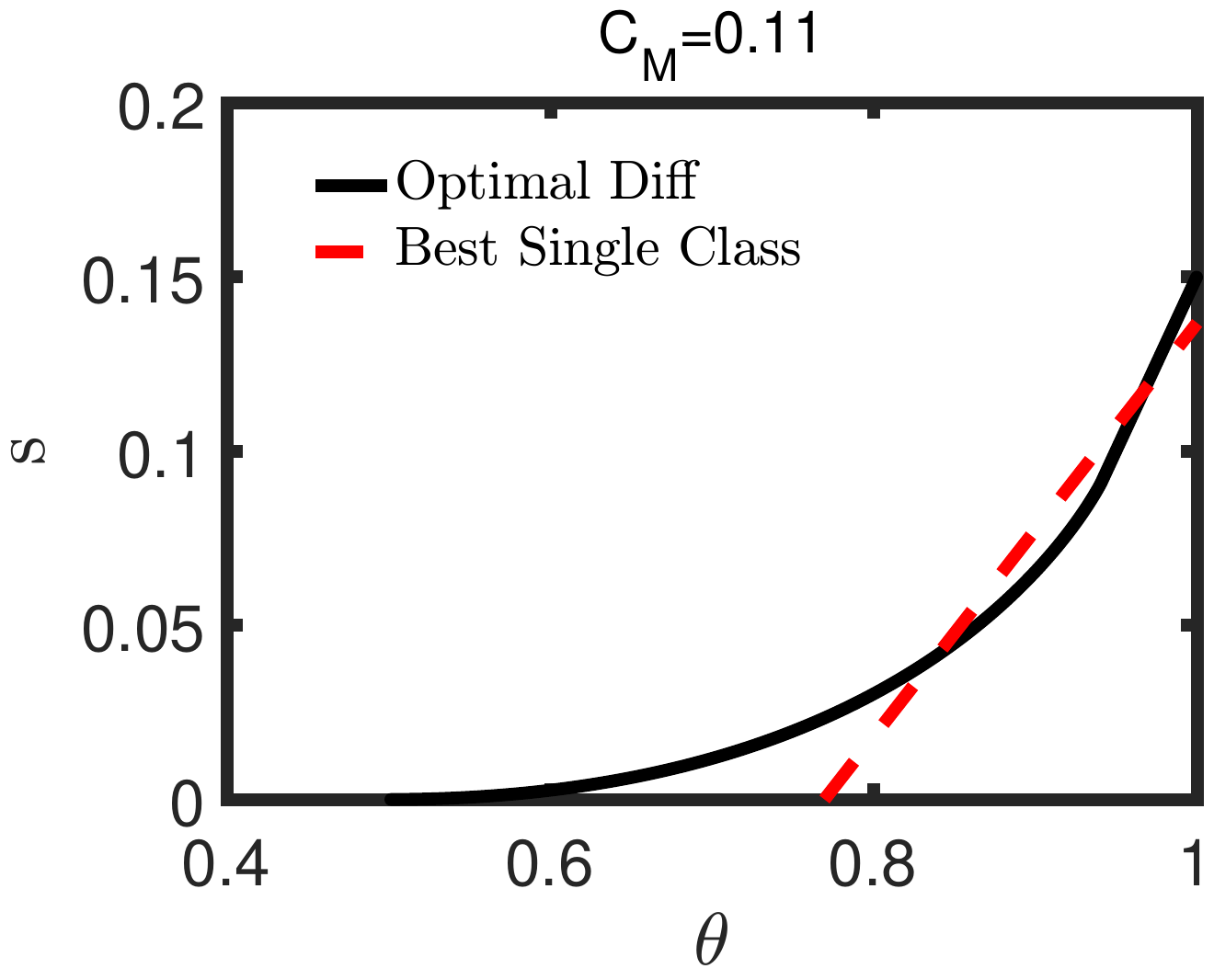}
\includegraphics[width=0.24\textwidth]{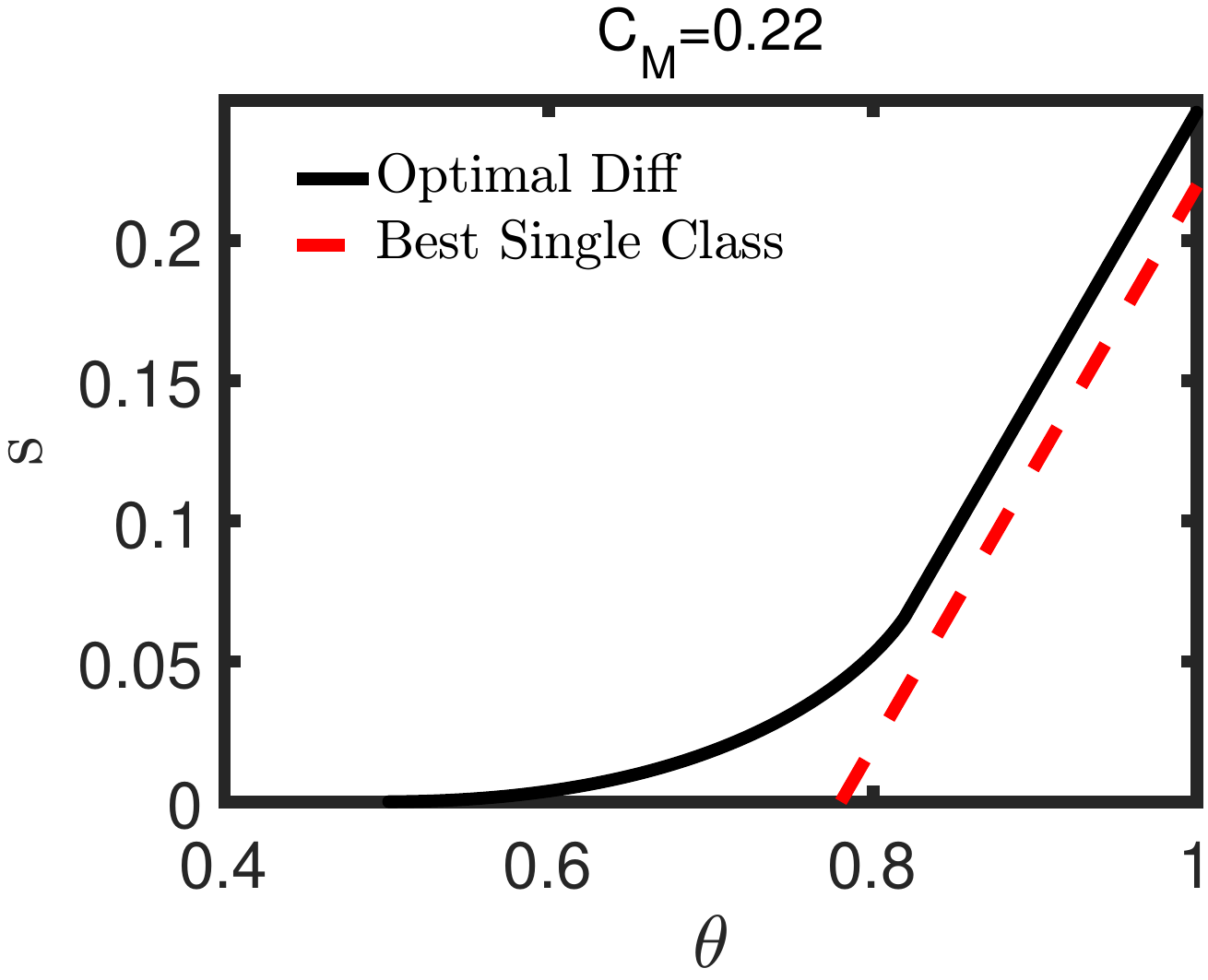}
\vspace{-2em}
\caption{Surplus of each user}
\vspace{-1em}
\label{single_vs_opt_surplus}
\end{figure}

The trend of total user surplus $s$ is similar, except for that the increase is also marginal when capacity is scarce. To see this more clearly, we plot the user surplus $s$ versus the user type $\theta$ when $C_M = 0.11$ (on the left) and $C_M = 0.22$ (on the right) in Figure \ref{single_vs_opt_surplus}.
We observe that under the ISP's optimal service differentiation, more users subscribe to the services.
However, when the capacity is insufficient, mid-tier uers are sacrificed  to subsidize the high-end and low-end users.

\textbf{Summary of implications:} The ISP has strong incentives to implement service differentiation if its network capacity is scarce and under this scenario, regulators should encourage such practices because more users will have access to the services and the total user surplus is higher.

\section{conclusion}
In this work, we study an ISP's optimal service differentiation strategy subject to its network capacity constraints in a congested network market. In particular, we consider two typical scenarios: the fixed capacity and  variable capacity scenarios. By incorporating the dual approach in optimal taxation theory \cite{mirrlees1971exploration}, we establish an optimal control framework to analyze the ISP's profit-maximizing problem. We derive the analytical solution to the ISP's optimal service differentiation and characterize its structures. Our results show that the ISP's optimal service differentiation scheme offers service classes with sufficiently high qualities and prices such that users with low value will not use the services, even when the network capacity is abundant. When the ISP expands its capacity, its market share will be increased by offering more premium service classes and reducing prices. Furthermore, when there exist more high-value users, the ISP should focus on them by offering a narrower range of premium service classes and charging higher prices. Our results also show that the ISP has strong incentives to implement service differentiation when its capacity is scarce and suggest that  regulators should encourage such practices as they increase the total user surplus. 
%As a direction of future work, we will explore the optimal service differentiation under competition.

\section*{Acknowledgment}
This work is supported by National Science Foundation of China (61379038), the Huawei Innovation Research Program (HIRP) and Ministry of Education of Singapore AcRF grant R-252-000-572-112.

\begin{appendix}
\subsection{Technical Proofs}
\begin{IEEEproof}[Proof of Lemma \ref{delimeter}]
If $1\geq \theta_1 \geq ... \geq \theta_{|\mathcal{N}|}$, we define $|\mathcal{N}|$ intervals as follows:
\begin{displaymath}
I_i \triangleq [\theta_i,1], \quad \forall i \in \mathcal{N}.
\end{displaymath}
Note that for $i=1,2,...,|\mathcal{N}|-1$,
\begin{align}
&\theta \in I_i   \iff \theta \geq \theta_i=\frac{p_i-p_{i+1}}{v(q_i)-v(q_{i+1})}\nonumber\\
&\iff \theta v(q_i)-p_i \geq \theta v(q_{i+1})-p_{i+1}\nonumber\\
&\iff \theta \ \ \text{prefers} \ \ \text{class} \ \ i \ \ \text{to} \ \ \text{class} \ \ i+1\nonumber
\end{align}
and for $i=|\mathcal{N}|$, $\theta \in I_i \iff \theta$ subscribes to the ISP. Therefore, we have
\begin{displaymath}
\theta \in [\theta_i,\theta_{i-1}] \iff \theta \in I_{\mathcal{N}}\cap ... \cap I_i \cap \bar I_{i-1} \cap ... \cap \bar I_1,
\end{displaymath}
which means $\theta$ prefers class $i$ to $i+1$, $i+1$ to $i+2$, ..., $\mathcal{N}$ to opting-out, and $i$ to $i-1$, ..., $2$ to $1$. Hence we have
\begin{displaymath}
\Theta_{i}(\mathbf{p},\mathbf{q})=[\theta_{i},\theta_{i-1}], \quad i=1,2,...,|\mathcal{N}|.
\end{displaymath}
\end{IEEEproof}

\begin{IEEEproof}[Proof of Lemma \ref{monotonicity}]
Suppose $q(\cdot)$ is not nonincreasing in $\theta$, then there exist two user types $\theta_1$ and $\theta_2$ such that $\theta_1>\theta_2$ and $q(\theta_1)>q(\theta_2)$. Let $q_1=q(\theta_1)$, $q_2=q(\theta_2)$, $p_1=p(q_1)$ and $p_2=p(q_2)$. Since the pricing function $p(\cdot)$ is strictly decreasing, we have $p_1<p_2$. By the definition of user's choice function $q(\cdot)$, we have the following two incentive inequalities:
\begin{align*}
&\theta_1 v(q_1)-p_1 \geq \theta_1 v(q_2)-p_2,\\
&\theta_2 v(q_2)-p_2 \geq \theta_2 v(q_1)-p_1.
\end{align*}
And they are equivalent to:
\begin{align*}
\theta_1 \leq \frac{p_2-p_1}{v(q_2)-v(q_1)}, \\
\theta_2 \geq \frac{p_2-p_1}{v(q_2)-v(q_1)}.
\end{align*}
which contradicts with that $\theta_1>\theta_2$.
\end{IEEEproof}

\begin{IEEEproof}[Proof of Lemma \ref{dual_approach}]
Since the utility function (\ref{utility}) has the \textit{strict and smooth single crossing differences} properties \cite{milgrom2004putting}, we can apply the constraint simplification theorem (see Theorem 4.3 in \cite{milgrom2004putting}) to derive this lemma.
\end{IEEEproof}

\begin{IEEEproof}[Proof of Lemma \ref{v0}]
Since $u(p,q;\theta=0)=-p\leq 0$ and $u(0,1;\theta=0)=0$ (choosing the dummy service class), we have $V^*(0)$=0. 
%\footnote{We remark that Lemma \ref{v0} actually holds for all $\theta \in [0,\hat\theta]$, which immediately follows Theorem \ref{fixedc_sol}.}
\end{IEEEproof}

\begin{IEEEproof}[Proof of Lemma \ref{capacity}]
Define the set of users that choose service class $q$ as
\begin{displaymath}
\Phi(q) \triangleq \{\theta \in \Theta : q(\theta)=q\}.
\end{displaymath}
We perform summation in a user-wise fashion instead of the class-wise one:
\begin{align*}
\int_\mathcal{Q}\big(\int_{\Phi(q)}f(\theta)\mathrm{d}\theta\big)w(q)\mathrm{d}q
&=\int_\mathcal{Q}\big(\int_{\Phi(q)}f(\theta)w(q)\mathrm{d}\theta\big)\mathrm{d}q\\
&=\int_\mathcal{Q}\big(\int_{\Phi(q)}f(\theta)w(q(\theta))\mathrm{d}\theta\big)\mathrm{d}q\\
%&=\int_\mathcal{Q}\big(\int_{\Phi(q)}f(\theta)\mathrm{d}\theta\big)w(q(\theta))\mathrm{d}q\\
&=\int_{0}^{1}w(q(\theta))f(\theta)\mathrm{d}\theta
\end{align*}
\end{IEEEproof}

\begin{IEEEproof}[Proof of Lemma \ref{fixed_control}]
The envelop theorem \cite{milgrom2002envelope} together with Lemma \ref{dual_approach} have established the one-one correspondence between the any service differentiation strategy and the corresponding user choice function $q(\cdot)$ together with an indirect utility level $V(0)$. From Lemma \ref{v0} we have set $V(0)=0$. From (\ref{W_def}) we obtain $W'(\theta)=w(q(\theta))f(\theta)$ and the total capacity consumption is $W(1)$ by Lemma \ref{capacity}. The first order derivative of $V(\cdot)$ is given by the envelop theorem. The ISP's profits is
\begin{align*}
\int_{\mathcal{Q}} p(q)\left(\int_{\Phi(q)}f(\theta)\mathrm{d}\theta\right) \mathrm{d}q
&=\int_{0}^{1} p(q(\theta))f(\theta)\mathrm{d}\theta\nonumber \\
&=\int_{0}^{1} [\theta v(q(\theta))-V(\theta)]f(\theta)\mathrm{d}\theta\nonumber
\end{align*}
Then Lemma \ref{fixed_control} is a summarization of the above results.
\end{IEEEproof}

\begin{IEEEproof}[Proof of Theorem \ref{fixedc_sol}]
In Section IV-C, we have introduced the three steps of tackling the optimal control problem (\ref{fixed_def}). In particular, we have proved that the candidate optimal control $\widetilde q(\cdot) $ for the relaxed optimal control problem is the point-wise maxima of
\begin{displaymath}
\Psi(\theta,q)\triangleq\left(\theta-\frac{1-F(\theta)}{f(\theta)}\right)v(q)-\mu \cdot w(q).
\end{displaymath}
The shadow price $\mu$ satisfies the slackness condition:
\begin{displaymath}
\mu \geq 0, \quad \mu\cdot[W(1)-C_M]=0
\end{displaymath}

If $W(1)-C_M<0$, then we have $\mu=0$, $\widetilde q(\cdot)$ is the point-wise maxima of
\begin{displaymath}
\Psi(\theta,q)=G(\theta)\cdot v(q)
\end{displaymath}
Under Assumption \ref{G}, $G(\theta)$ is negative on the interval $[0,\theta_0)$ and positive on the interval $(\theta_0,1]$. Since $v(\cdot)$ is decreasing in $q$, it naturally follows that $\widetilde q(\cdot)=1$ for all $\theta \in [0,\theta_0)$ and $\widetilde q(\cdot)=0$ for all $\theta \in [\theta_0,1]$. Then we have the second part of Theorem \ref{fixedc_sol}. Moreover, we have obtained the \textit{abundance capacity} $\bar C \triangleq [1-F(\theta_0)]w(0)$, the corresponding capacity consumption level.

If $W(1)-C_M=0$, the slackness condition gives $\mu>0$. Then $\widetilde q(\cdot)$ is the point-wise maxima of
\begin{displaymath}
\Psi(\theta,q)=G(\theta)\cdot v(q) -\mu \cdot w(q).
\end{displaymath}
When $\theta \in [0,\theta_0)$, $G(\theta)<0$ and $\Psi(\theta,q)$ is decreasing in $q$. Therefore $\widetilde q(\cdot)=1$ for all $\theta \in [0,\theta_0)$. When $\theta > \theta_0$, we calculate the partial derivative with respect to $q$ to find the maxima.
\begin{displaymath}
\frac{\partial \Psi(\theta,q)}{\partial q}=G(\theta)v'(q)-\mu \cdot w'(q)=-\mu \cdot v'(q)[h(q)-G(\theta)/\mu]
\end{displaymath}
Since $-\mu \cdot v'(q)>0$, the sign of $\frac{\partial \Psi(\theta,q)}{\partial q}$ is determined by $h(q)-G(\theta)/\mu$. Then under Assumption \ref{h}, we have
\[ \widetilde q(\cdot)=\begin{cases}
      1, &\theta \in [0,\hat\theta) \\
      h^{-1}(G(\theta)/\mu), & \theta \in [\hat\theta,1]; \\
   \end{cases}
\]
where the marginal user is $\hat\theta=G^{-1}(\mu \cdot h(1))$.

By far we have completed the first step, i.e. to solve the candidate optimal control $\widetilde q(\cdot)$ for the relaxed optimal control problem. Next we prove that $\widetilde q(\cdot)$ is indeed the optimal control of the relaxed optimal control problem. This is guaranteed by the Arrow sufficiency theorem \cite{chiang2000elements} due to that the Hamiltonian is concave in $V$. As for the third step, it is easy to show that $\widetilde q(\cdot)$ is nonincreasing under Assumption \ref{h}. Hence we have $q^*(\cdot)=\widetilde q(\cdot)$ and the proof of Theorem \ref{fixedc_sol} is completed.

\end{IEEEproof}

\begin{IEEEproof}[Proof of Theorem \ref{expansion}]
For the variable capacity scenario, the transversality conditions become
\begin{align*}
\begin{cases}
\displaystyle\frac{d\lambda_1}{d\theta}=-\frac{\partial H}{\partial V}=f(\theta)\vspace{0.05in}\\
\displaystyle\frac{d\lambda_2}{d\theta}=-\frac{\partial H}{\partial W}=0\\
\displaystyle\lambda_1(1)=0, \quad \lambda_2(1)=-S'(\widetilde W(1)).
\end{cases}
\end{align*}

Therefore, the costate variable associated with $W(\cdot)$ is a constant $\lambda_2(\theta)\equiv -S'(\widetilde W(1))$. In parallel with Theorem 1, the marginal cost $S'(\widetilde W(1))$ at the optimal capacity consumption level $\widetilde W(1)$ here plays the same role as the shadow price $\mu$ in the fixed capacity scenario. Assuming $S(\cdot)$ is nondecreasing and convex, the three steps in the proof of Theorem \ref{fixedc_sol} can be readily applied in this scenario and we omit the details here.
\end{IEEEproof}

\begin{IEEEproof}[Proof of Corollary \ref{optimalstrategy}]
Note that the indirect utility level is $V^*(0)=0$, then Corollary \ref{optimalstrategy} is straightforward from Lemma \ref{dual_approach}.
\end{IEEEproof}

\begin{IEEEproof}[Proof of Corollary \ref{shadow}]
By Theorem \ref{fixedc_sol}, If $C_M < \bar C$, then the user choice function $q^*(\cdot)$ and the shadow price $\mu$ are uniquely determined by the following set of equations.\\
\textit{a)} The user choice function:
\[ q^*(\theta)=\begin{cases}
      1, &\theta \in [0,\hat\theta) \\
      h^{-1}(G(\theta)/\mu), & \theta \in [\hat\theta,1]; \\
   \end{cases}
\]
\textit{b)}
\begin{displaymath}
\hat\theta=G^{-1}(\mu \cdot h(1));
\end{displaymath}
\textit{c)} The capacity equation:
\begin{displaymath}
\int\nolimits_{0}^{1}w(q^*(\theta))f(\theta)\mathrm{d}\theta = C_{M}.
\end{displaymath}

When $F(\cdot)$ is fixed, $C_M$ is a function of $\mu$. Suppose the shadow prices under two maximum capacities $C_1$ and $C_2$ are $\mu_1$ and $\mu_2$ respectively and $0< \mu_1 < \mu_2$. From the capacity equations, we have
\begin{align*}
C_1&=\int_{0}^{1}w(q_1^*(\theta))f(\theta)\mathrm{d}\theta=\int_{\hat\theta_1}^{1}w(q_1^*(\theta))f(\theta)\mathrm{d}\theta\\
C_2&=\int_{0}^{1}w(q_2^*(\theta))f(\theta)\mathrm{d}\theta=\int_{\hat\theta_2}^{1}w(q_2^*(\theta))f(\theta)\mathrm{d}\theta\\
\end{align*}
From Theorem \ref{fixedc_sol} we have $\hat\theta_1<=\hat\theta_2$. Moreover, if $h(0)=+\infty$, then $q_1(\theta)<q_2(\theta)$ for all $\theta \in [\hat\theta_2,1]$; if $h(0)<+\infty$, letting $\bar\theta_2 \triangleq \min\{G^{-1}(\mu_2 \cdot h(1)),1\}$, then $q_1(\theta)<q_2(\theta)$ for all $\theta \in [\hat\theta_2, \bar\theta_2]$. Therefore, we obtain $C_1 > C_2$. Furthermore, we have $\lim_{\mu \to 0}C=\bar C$ . By far we have established the one-one correspondence between $C_M$ and $\mu$ and thus indirectly proving Corollary \ref{shadow}.
\end{IEEEproof}

\begin{IEEEproof}[Proof of Corollary \ref{exclude}]
Given Theorem \ref{fixedc_sol} and Corollary \ref{shadow}, this result is straightforward.
\end{IEEEproof}

\begin{IEEEproof}[Proof of Corollary \ref{bunching}]
Actually $\hat C$ is such that the corresponding shadow price $\hat\mu=G(1)/h(0)=1/h(0)$. When $C_M=\hat C$, user with the highest value $\theta=1$ chooses the best service with congestion $q=0$. The rest of the proof is straightforward.
\end{IEEEproof}

\begin{IEEEproof}[Proof of Corollary \ref{capacity_theorem}]
The range of congestion levels $\mathcal{Q}^*=[q^*(1),q^*(\hat\theta)]=[h^{-1}(1/\mu),1]$. From Corollary \ref{shadow}, we have proved the first part of the corollary that if $C_{M1}<C_{M2}$, then $\mathcal{Q}_{1} \subset \mathcal{Q}_{2}$.
%For the second part, we consider how the price $p(q)$ changes with $C_M$. We focus on the following identity:
%\begin{equation}\label{pricing_identity}
%p(q(\theta))=\theta v(q(\theta))-V(\theta).
%\end{equation}
%For $q=0 \in \mathcal{Q}$, we have $q(1)=0$. Letting $\theta=1$ in (\ref{pricing_identity}) we have $p(0)=1-V(1)=1-\int_0^1 v(q(\theta))\mathrm{d}\theta$. By Corollary \ref{shadow}, it is easy to see $p(0)$ deceases with $C_M$.
\end{IEEEproof}

\begin{IEEEproof}[Proof of Corollary \ref{profits}]
The ISP's optimal profits is
\begin{displaymath}
J^*=\int_0^1 [\theta q^*(\theta) -V^*(\theta)]f(\theta)\mathrm{d}\theta.
\end{displaymath}
Integrating by parts, we have
\begin{align*}
\int_0^1 V^*(\theta)f(\theta)\mathrm{d}\theta&=\int_0^1 V^*(\theta)\mathrm{d}F(\theta)\\
&=[V(\theta)F(\theta)]\mid_0^1-\int_0^1 F(\theta)dV^*(\theta)\\
&=V(1)-\int_0^1 F(\theta)v(q^*(\theta))\mathrm{d}\theta\\
&=\int_0^1 [1-F(\theta)]v(q^*(\theta))\mathrm{d}\theta.
\end{align*}

Hence we have
\begin{displaymath}
J^*=\int_0^1 G(\theta)v(q^*(\theta))f(\theta)\mathrm{d}\theta.
\end{displaymath}

By Corollary \ref{shadow}, we have that $J^*$ increases with $C_M<\bar C$ and the asymptotic limit is reached when $C_M=\bar C$.
\end{IEEEproof}

\subsection{Extension to Pay-As-You-Go Pricing}
We briefly demonstrate how our analysis can be extended to handle pay-as-you-go pricing scheme. We now interpret $v(q)$ as users' volume of transmitted packets under congestion level $q$. The user with value $\theta$ has utility $v(q)(\theta - p)$ when joining the class with congestion $q$ and price $p$. If we define the indirect utility and user choice as follows.
The \textit{indirect utility} of user type $\theta$ is given by
\begin{displaymath}
V(\theta) \triangleq \max_{q\in\mathcal{Q}} \  v(q)(\theta - p)
\end{displaymath}
The user type $\theta$ must subscribe to a service class in the set
\begin{displaymath}
X(\theta) \triangleq \arg\max_{q\in\mathcal{Q}}  v(q)(\theta - p)
\end{displaymath}
i.e., the choice of user $\theta$ is given by \(q(\theta) \in X(\theta)\).

Then we have $V'(\theta)=v(q(\theta))$, and $p(q(\theta))=\theta-V(\theta)/v(q)$. The ISP's profits is
\begin{displaymath}
\int_0^1 p(q(\theta))v(q(\theta))f(\theta)\mathrm{d}\theta=\int_0^1 [\theta v(q(\theta))-V(\theta)]f(\theta)\mathrm{d}\theta.
\end{displaymath}

In the fixed capacity situation, the optimal control problem is
\begin{align*}
\underset{q(\cdot)}{\text{maximize}}\quad &J=\int_{0}^{\theta_{M}} [\theta v(q(\theta))-V(\theta)]f(\theta)\mathrm{d}\theta \\
\text{subject to}\quad &V'(\theta)=v(q(\theta)),\quad V(0)=0 \\
&W'(\theta)=w(q(\theta))v(q(\theta))f(\theta),\quad W(0)=0 \\
&W(\theta_{M})\leq C_{M} \\
&q(\cdot)\quad\text{nonincreasing},\quad 0 \leq q(\theta) \leq 1
\end{align*}

Comparing the above optimal control problem with (\ref{fixed_def}), we see that in the pay-as-you-go scheme, $w(\cdot)v(\cdot)$ plays the same role as $w(\cdot)$ in the flat-rate scheme. Therefore our analysis can be extended to handle the pay-as-you-go scheme with easy modifications.

\end{appendix}

%p
%\bibliographystyle{abbrv}
%\bibliography{sigproc}  % sigproc.bib is the name of the Bibliography in this case
% You must have a propper ".bib" file
% ACM needs 'a single self-contained file'!
%\balancecolumns
{
%\begin{small}
\bibliographystyle{abbrv}
\bibliography{paper}
%\end{small}
}

\end{document}